\documentclass[prx, superscriptaddress,showkeys]{revtex4-1}
\usepackage{titlesec}
\usepackage[version=4]{mhchem}
\usepackage{graphicx}% Include figure files
\usepackage{dcolumn}% Align table columns on decimal point
\usepackage{bm}% bold math
\usepackage{color}
\usepackage{tabularx}
\usepackage{tabularray}
\usepackage{threeparttable}
\usepackage{array}
\usepackage{amsmath}
\usepackage{amssymb}
\usepackage{mhchem}
\usepackage{comment}
\usepackage{pbox}
\usepackage{float}
\usepackage[utf8]{inputenc}
\DeclareUnicodeCharacter{0394}{$\Delta$}

\usepackage{stmaryrd}  %\llbracket \rrbracket

\newcommand{\ignore}[1]{}

\begin{document}

\preprint{APS/123-QED}

\title{Spin Dynamics in van der Waals Magnetic Systems}% Force line breaks with \\

\author{Chunli Tang}
\affiliation{Department of Electrical and Computer Engineering, Auburn University, Auburn, AL 36849, USA}
\affiliation{Department of Physics, Auburn University, Auburn, AL 36849, USA}

\author{Laith Alahmed}
\affiliation{Department of Electrical and Computer Engineering, Auburn University, Auburn, AL 36849, USA}

\author{Muntasir Mahdi}
\affiliation{Department of Electrical and Computer Engineering, Auburn University, Auburn, AL 36849, USA}

\author{Yuzan Xiong}
\affiliation{Department of Physics, Oakland University, Rochester, MI 48309 USA}

\author{Jerad Inman}
\affiliation{Department of Physics, Oakland University, Rochester, MI 48309 USA}

\author{Nathan J. McLaughlin}
\affiliation{Department of Physics, University of California, San Diego, La Jolla, California 92093, USA}

\author{Christoph Zollitsch}
\affiliation{London Centre for Nanotechnology, University College London, 17-19 Gordon Street, London, WCH1 0AH, UK}

\author{Tae Hee Kim}
\affiliation{Center for Quantum Nanoscience, Institute for Basic Science (IBS), Seoul 03760, South Korea}
\affiliation{Department of Physics, Ewha Womans University, Seoul 03760, South Korea}

\author{Chunhui Rita Du}
\affiliation{Department of Physics, University of California, San Diego, La Jolla, California 92093, USA}

\author{Hidekazu Kurebayashi}
\affiliation{London Centre for Nanotechnology, University College London, 17-19 Gordon Street, London, WCH1 0AH, UK}
\affiliation{Department of Electronic and Electrical Engineering, University College London, Roberts Building, London, WC1E 7JE, United Kingdom}
\affiliation{WPI Advanced Institute for Materials Research, Tohoku University, 2-1-1, Katahira, Sendai 980-8577, Japan}

\author{Elton J. G. Santos}
\email{esantos@ed.ac.uk}
\affiliation{Institute for Condensed Matter Physics and Complex Systems, School of Physics and Astronomy, The University of Edinburgh, Edinburgh, EH9 3FD, UK}
\affiliation{Higgs Centre for Theoretical Physics, The University of Edinburgh,  EH9 3FD, UK}

\author{Wei Zhang}
\email{zhwei@unc.edu}
\affiliation{Department of Physics and Astronomy, University of North Carolina at Chapel Hill, NC 27599, USA}
\affiliation{Department of Physics, Oakland University, Rochester, MI 48309 USA}

\author{Peng Li}
\email{lipeng18@ustc.edu.cn}
\affiliation{School of Microelectronics, University of Science and Technology of China, Hefei 230052, China}
\affiliation{Department of Electrical and Computer Engineering, Auburn University, Auburn, AL 36849, USA}

\author{Wencan Jin}
\email{wjin@auburn.edu}
\affiliation{Department of Physics, Auburn University, Auburn, AL 36849, USA}
\affiliation{Department of Electrical and Computer Engineering, Auburn University, Auburn, AL 36849, USA}

\date{\today}% It is always \today, today,
             %  but any date may be explicitly specified

\begin{abstract}
The discovery of atomic monolayer magnetic materials has stimulated intense research activities in the two-dimensional (2D) van der Waals (vdW) materials community. The field is growing rapidly and there has been a large class of 2D vdW magnetic compounds with unique properties, which provides an ideal platform to study magnetism in the atomically thin limit. In parallel, based on tunneling magnetoresistance and magneto-optical effect in 2D vdW magnets and their heterostructures, emerging concepts of spintronic and optoelectronic applications such as spin tunnel field-effect transistors and spin-filtering devices are explored. While the magnetic ground state has been extensively investigated, reliable characterization and control of spin dynamics play a crucial role in designing ultrafast spintronic devices. Ferromagnetic resonance (FMR) allows direct measurements of magnetic excitations, which provides insight into the key parameters of magnetic properties such as exchange interaction, magnetic anisotropy, gyromagnetic ratio, spin-orbit coupling, damping rate, and domain structure. In this review article, we present an overview of the essential progress in probing spin dynamics of 2D vdW magnets using FMR techniques. Given the dynamic nature of this field, we focus mainly on broadband FMR, optical FMR, and spin-torque FMR, and their applications in studying prototypical 2D vdW magnets. We conclude with the recent advances in laboratory- and synchrotron-based FMR techniques and their opportunities to broaden the horizon of research pathways into atomically thin magnets. 

\end{abstract}

%Use showkeys class option if keyword
                              %display desired
\maketitle

\tableofcontents

\section{Introduction}

\begin{figure*}   
\centering
\includegraphics[width=0.90\textwidth]{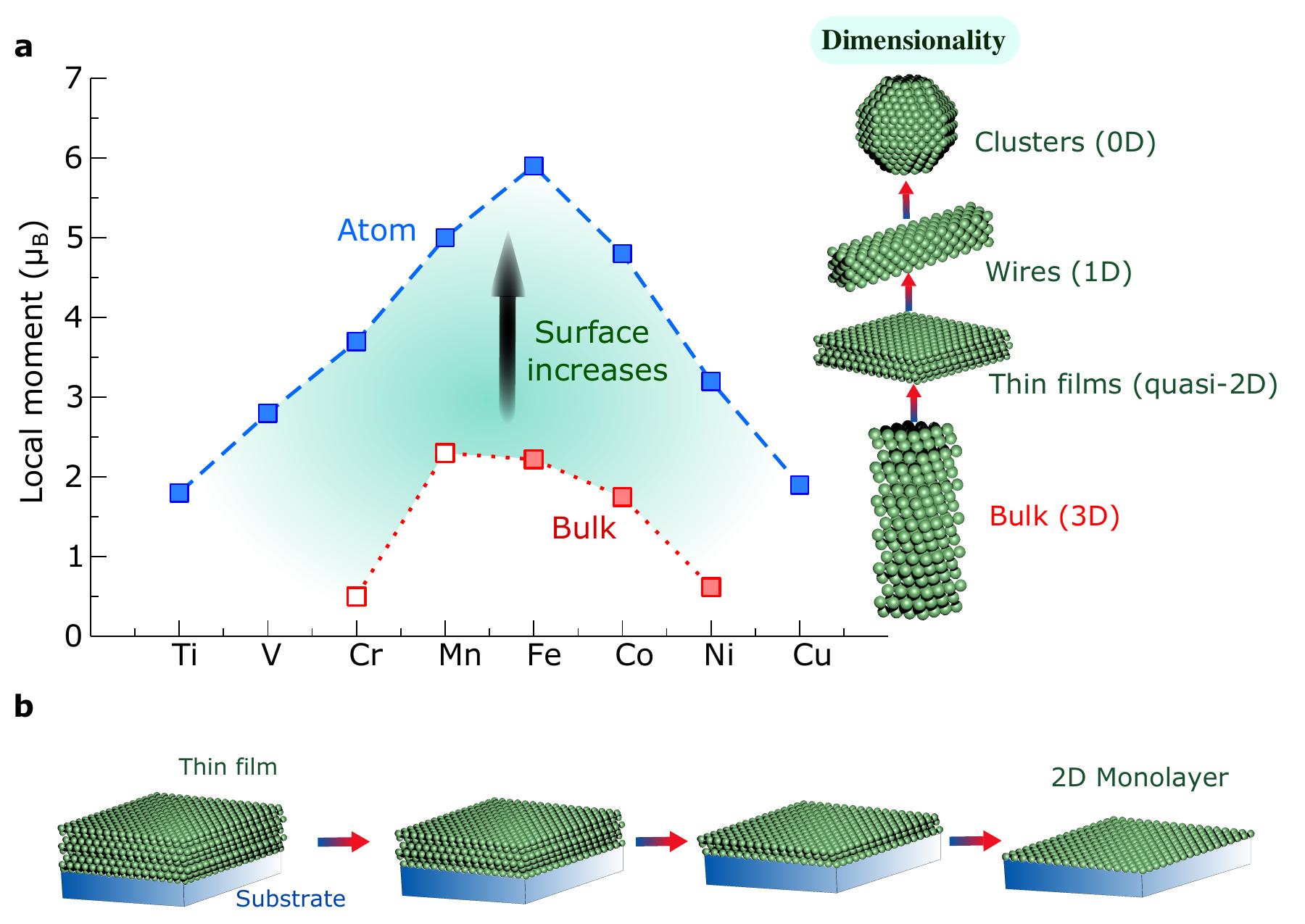}
\caption{\label{dimensionality} \textbf{Dimensionality induced variation of magnetic properties.} 
\textbf{a.} Local magnetic moments of several 3$d$ transition metals at bulk and at the isolated atom. Only $d$-electrons are considered for the atoms whereas the spin moment is included for bulk~\cite{Coey-book}. Ferromagnetic (antiferromagnetic) bulk metals are shown by filled (empty) squares. 
{\bf b.} Schematic of a thin film deposited on top of a substrate with decreasing thickness at specific growth conditions until reaches the 2D monolayer limit. 
}
\end{figure*}

The demonstration of magnetism in the two-dimensional (2D) limit has long been the focus of fundamental questions in condensed matter physics ~\cite{onsager1944crystal}. In the 1960s, quasi-2D magnetic order has been identified in bulk magnets in which the weak interlayer interactions prevent magnetic ordering in the third dimension ~\cite{plumier1964neutron, lines1967examples, lines1969magnetism}. In the 1970s, ultrathin film magnetism was realized in elemental metals or metallic alloys thanks to the development of vacuum-deposition techniques ~\cite{gradmann1968very, gradmann1974ferromagnetism, gradmann1985magnetic}. 
%%%%
Such approaches allow to deposit layers of a magnetic compound atom-by-atom on a substrate via vapor deposition techniques, i.e., chemical vapor deposition (CVD) or physical vapor deposition (PVD). That is, the atoms or molecules from a vapor source reach the solid-state surface with almost no collision with residual gas molecules in an inert environment (e.g., deposition chamber), which chemical reactions might take place at the surface to promote the formation of the system~\cite{vapor-book}. Different precursors, rate diffusion, annealing, pressure, and temperature conditions are used to induce the formation of high-crystalline samples. Good vacuum is normally required for the processes involved as the deposited layers can range from one-atom thickness up to millimeters. PVD and CVD allow the fabrication of high-quality, high-purity, highly controlled elemental and permalloy thin films, but suffer to be able to produce layers that can be easily exfoliated as in the case of 2D van der Waals (vdW) materials. Indeed, transition-metal monolayers (e.g., V, Cr, Mn, Fe, Co, Ni)  deposited on different substrates (e.g., noble-metals)~\cite{Jonas92, Walker97, Fitzsimmons98, Arabski94, Diebold97, Bucher98, Blugel01, Hoving89, DeVries96, Bloemen91} were considered historically classical systems exhibiting 2D magnetism. The reduced dimensionality induces a low coordination number of nearest neighbor atoms for the transition metal which intrinsically changes several electronic features relative to bulk. Such as {\it d-}band widths, local density of states at the Fermi level, magnetic moment and charge density~\cite{Blugel89}. For instance, the large variation of the magnetic moment from bulk down to the atom limit provides a number of nanoscale systems (e.g., clusters, wires, thin films) to be explored (Figure \ref{dimensionality}{\bf a}). At the atom limit, the magnetic moment follows Hund's first rule resulting in almost third transition-metal elements with a net magnetic moment over the entire periodic table. This counting argument based on the electron filling of the electronic shells indicates that most of the large magnitudes of the {\it 3d} moments are located at the middle of the series around Cr and Mn (5 $\mu_B$). This scenario changes critically at bulk as most of the transition metal compounds do not show any magnetic features with the exception of five elements: Cr, Mn, Fe, Co, and Ni. Between single atoms and bulk crystals, a broad amount of physical behavior determines the magnetic characteristics~\cite{Whitehead00} of the systems.

In this context, the {\it d-d} hybridization between the transition metal monolayer and a metallic substrate (Figure \ref{dimensionality}{\bf b}) is an additional parameter to be considered as it affects the magnetic properties of the layers as reported in many systems~\cite{Bloemen91,DeVries96}. For instance, magnitudes of the magnetic moment~\cite{Blugel00V}, ground states (ferromagnetic, or anti-ferromagnetic)~\cite{Wiesendanger00_sc}, formation of frustrated spin-ordered~\cite{Wiesendanger00,Marcus92}, interlayer spin relaxation~\cite{Gauthier93}, magnetocrystalline anisotropy~\cite{Hoving89,Bruno89}, and exchange energies~\cite{Abt00,Whitehead00} have been observed to show some dependence with the underneath surfaces. Initially assumed to be a tuning parameter for magnetic monolayers deposited on noble metal surfaces, interfacial interactions promptly became more complex to deal chemically which inherently obscured more fundamental phenomena thought to be present in strictly two dimensions. As a matter of fact, the long-time until truly monolayer materials could be isolated or transferred to different substrates for a variety of applications might be due to the common belief that interactions with the substrates are strong enough to harden any possibility of free-standing magnetic layers absent of any substrates. This scenario has proved to be true until early 2000s when the first works on 2D van der Waals (vdW) materials came to light~\cite{Geim05,Firsov05,Kim05}.

Nonetheless the past research in quasi-2D magnets and magnetic thin films resulted in unfruitful developments, since a wide variety of spintronic functionality (i.e., magnetic tunnel junctions and giant magnetoresistance ~\cite{julliere1975tunneling, baibich1988giant, binasch1989enhanced, dieny1991giant, dieny1994giant, moodera1999spin, zhu2006magnetic}) has been exploited in data storage and random-access memory applications using underlying principles from quasi-2D systems. Those developments have allowed a revolution on how thin, fast, and low-cost data information can become. There have been around 12 years since the isolation of graphene and other 2D materials~\cite{Geim05} until on 2017 when intrinsic 2D magnetic order was discovered in vdW Cr$_2$Ge$_2$Te$_6$ ~\cite{gong2017discovery} and CrI$_3$ ~\cite{huang2017layer} crystals down to atomically thin flakes. The success of this approach highlights that the intrinsic character of short-range interactions in finite systems can overcome thermal fluctuations and stabilize long-range magnetic order in the 2D limit~\cite{Jenkins22}. 

Despite the new opportunities presented by 2D magnetism, the experimental techniques used to study 2D magnets are still less mature than those used for the well-established bulk magnets and magnetic thin films. Since 2D vdW magnets are typically produced as micrometer-scale thin flakes, macroscopic volumetric analysis tools such as superconducting quantum interference devices (SQUID) magnetometry and neutron scattering are generally not applicable. Optical spectroscopy and microscopy such as polarized micro-Raman scattering ~\cite{wang2016raman, jin2018raman, zhang2019magnetic, kim2019raman, mccreary2020distinct, mccreary2020quasi, huang2020tuning, li2020magnetic, guo2021structural, sun2021review, wang2022electronic}, magneto-optical Kerr effect (MOKE) ~\cite{huang2017layer, huang2018electrical} and magnetic circular dichroism (MCD) ~\cite{jiang2018electric, seyler2018ligand} have been widely used to study magnetic order parameters of 2D magnets. However, these techniques are highly sensitive to the wavelength of light and the choice of substrate since the magneto-optical response, as shown in CrI$_3$ and CrBr$_3$, is dominated by excitonic effect ~\cite{wu2019physical, jin2020observation, wu2022optical}.

Ferromagnetic resonance (FMR) has been broadly applied to a range of materials from bulk magnets to nano-scale magnetic thin films. There are several excellent reviews on this technique ~\cite{frait1988spin, heinrich1993ultrathin, farle1998ferromagnetic, maksymov2015broadband, von2016ferromagnetic, schmool2021ferromagnetic}. Nowadays, the unique capabilities of FMR have been exploited to study 2D magnets. First, standard FMR theory shows that the resonance frequency is a function of the effective field, which comprises the information of the exchange coupling, magneto-crystalline and shape anisotropy ~\cite{kittel1948theory, phillips1966spin, barnes1981theory}. This dependence can be used to characterize the microscopic parameters of 2D magnets. In addition, the dynamic properties of 2D magnets can be easily perturbed by microwave absorption. Therefore, FMR can be used to explore magnetization dynamics due to varying mechanisms including spin pumping, spin torque, and spin Hall effect. Moreover, analysis of FMR line shape provides direct access to the damping constant, which is the most important parameter for controlling dynamic behaviors in spintronic devices.

In this review, we summarize the recent development of FMR studies of 2D vdW magnets. The article is arranged into sections as follows. Section 1 provides an overview of 2D vdW magnets and highlights their unique characteristics in magnetization dynamics. Section 2 introduces the formalism of magnetization dynamics using standard FMR theory and prototypical FMR techniques. Section 3 shows the broadband FMR spectroscopy studies of the key parameters of magnetic properties including magnetic anisotropy, $g$-factor, domain structure, and interlayer coupling in the prototypical 2D vdW magnets. Section 4 discusses the optical excitation and detection of magnetization dynamics and the usage of this technique in studying laser-induced magnetization dynamics, magnetic critical behavior, light control of magnetism, and collective excitations such as exciton-magnon and phonon-magnon coupling. Section 5 presents the current induced spin-torque resonance and highlights its application in 2D-magnets-based spin-orbit torque devices. In Section 6, we discuss the recent technical advances in magnetization dynamics with phase, energy, spatial, and element resolution. Finally, we conclude with future perspectives and opportunities in this field.

%%%%%%%%%%%%%%%%%%%%%%%%%%%%%%%%%%%%%%%%%%%%%%%%%%%%%%%%%%%%%%%%%%%

\section{Overview of 2D vdW magnets}

Using mechanical exfoliation, chemical vapor deposition (CVD), and molecular beam epitaxy (MBE) methods, a rich collection of vdW magnetic materials covering a wide spectrum of magnetic properties are experimentally demonstrated including binary transition metal halides (VI$_3$ ~\cite{tian2019ferromagnetic_VI3}, CrCl$_3$ ~\cite{cai2019atomically}, CrBr$_3$ ~\cite{zhang2019direct_CrBr3}), binary transition metal chalcogenides (CrTe$_2$ ~\cite{sun2020room, meng2021anomalous, zhang2021room_CrTe2}, VSe$_2$ ~\cite{bonilla2018strong_VSe2}, VTe$_2$ and NbTe$_2$ ~\cite{li2018synthesis_VTe2}, Cr$_3$X$_4$ (X = S, Se, Te) ~\cite{zhang2019high_Cr3X4}), ternary transition metal compounds (MPS$_3$ or MPSe$_3$ (M = Mn, Fe, Ni) ~\cite{du2016weak}, Fe$_n$GeTe$_2$ (n = 3, 4, 5) ~\cite{deng2018gate, fei2018two, seo2020nearly, may2019ferromagnetism}, MnBi$_2$Te$_4$ ~\cite{deng2020quantum}, CrSBr ~\cite{jiang2018screening}, CrSiTe$_3$ ~\cite{li2022anomalous}), and other binary transition metal compounds (MnSe$_x$ ~\cite{o2018room}, CrB$_2$ ~\cite{park2020momentum}). For a more extended summary of currently known 2D vdW magnets and their synthesis methods, several recent review articles focused on the material landscape are available \cite{zhang2021two, jiang2021recent, xu2022recent,wang2022magnetic}. 

As the library of 2D magnets has been rapidly growing over the past few years, there are a couple of important research thrusts that attracted great attention in the field of materials science, microscopic theory, and device applications. First, a large variety of vdW magnets is available and many of their 2D form have not been investigated yet. The magnetic states in 2D materials are distinct from those in bulk crystals. For example, \ce{CrI3} in few-layer form develops layered antiferromagnetic (AFM) configuration below the magnetic onset of 45 K ~\cite{huang2017layer}, while bulk \ce{CrI3} exhibits a ferromagnetic (FM) order below the Curie temperature of 61 K ~\cite{mcguire2015coupling}. The layered AFM states give rise to giant tunneling magnetoresistance, which can be considered as a perfect spin filter ~\cite{song2018giant, klein2018probing, kim2018one, wang2018very}. The layered AFM states can be switched to FM states upon applying a moderate magnetic field ~\cite{huang2017layer}, or electric field ~\cite{huang2018electrical, jiang2018electric}, or electrostatic doping ~\cite{jiang2018controlling}, or hydrostatic pressure ~\cite{li2019pressure, song2019switching}, opening the route for the development of tunneling-based memory and sensing devices. Therefore, it is instructive to find more 2D semiconducting magnets with layered AFM configuration and to check whether the intimate coupling between spin configuration, electronic state, and stacking symmetry can exist in different compounds as it does in \ce{CrI3}. In addition, to realize industrial applications, the synthesis of 2D magnets with high transition temperature using wafer-scale methods is still pressingly in need. Second, to gain fundamental insights into the 2D magnetic order, it is crucial to understand the microscopic parameters such as dipolar interactions and magnetocrystalline anisotropy. Although the strength of these interactions is typically small, they will have a crucial impact on the 2D vdW magnets. It was initially thought that magnetic anisotropies~\cite{mermin1966absence} would be necessary for the stabilization of long-range ferromagnetic or antiferromagnetic order in 2D system at finite temperatures. However, recent results~\cite{Jenkins22} demonstrated that such dependence does not exist which opened the prospects of materials following a simple isotropic Heisenberg model. Third, in addition to the ground state, magnetic excitations provide a basic ingredient of the microscopic description of many physical properties. 
For example, magnons or spin waves are collective excitations of the spins in magnetically ordered systems. If their control via different driving forces (e.g., electrical currents, strain, laser pulses) can be achieved in real device-platforms intriguing energy-efficient applications might be created~\cite{wang2022magnetic}. Moreover, magnetic excitations play a key role in determining the spin pumping efficiency and switching speed of many spintronic devices. The recent progress of the field of 2D magnets has been extensively reviewed in the literature ~\cite{burch2018magnetism, gibertini2019magnetic, gong2019two, mak2019probing, cortie2020two, jiang2021recent, sierra2021van, och2021synthesis, yao2021recent, wang2022magnetic, kurebayashi2022magnetism}.

We also note some distinct characteristics where the 2D vdW systems are unique in comparison with the FMR studies of conventional 3D magnets and magnetic thin films. First, due to the small volume of samples, accurate characterization of magnetic properties of 2D magnets requires ultrahigh sensitivity. This makes several magnetization dynamics techniques very relevant to the study of 2D magnetism. For example, in determining saturation magnetization, magnetization dynamics is more advantageous than static magnetometry, as the magnetization is extracted from their intrinsic resonant dispersion relation (frequency vs. field), rather than picking up the total induction. Second, unlike bulk and magnetic multilayer systems, the 2D vdW systems concern only a few layers and the conventional thickness arguments become a purely stacking issue. The well-defined intralayer and interlayer couplings in 2D magnets thus underpin robust and reproducible spin dynamics that can be often modeled in the frame of synthetic (anti-)ferromagnetic systems. Third, many 2D vdW magnets are low-to-mid bandgap semiconductors. Therefore, exciton physics in the frame of optics can be simultaneously explored with its spin dynamics using ultrfast spectroscopy and magneto-optic effects.

%%%%%%%%%%%%%%%%%%%%%%%%%%%%%%%%%%%%%%%%%%%%%%%%%%%%%%%%%%%%%%%%%%%

\section{Magnetization dynamics and ferromagnetic resonance}

\subsection{Equation of magnetization dynamics and resonance condition}

\begin{figure*}   
\centering
\includegraphics[width=1.0\textwidth]{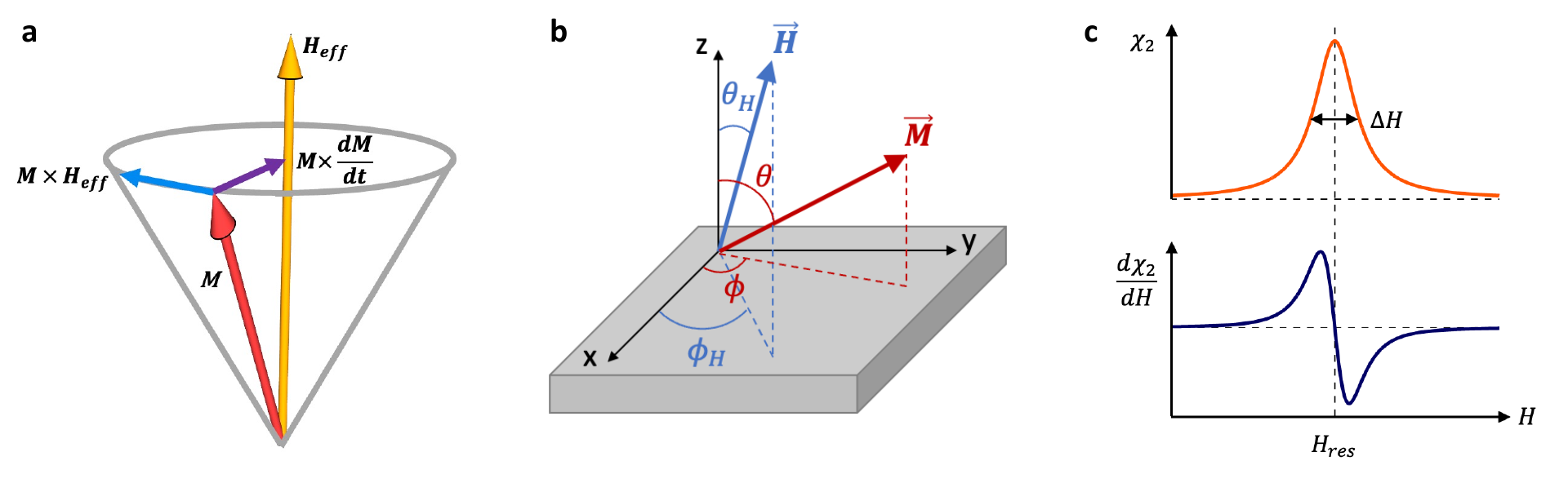}
\caption{\label{LLG} \textbf{Principles of magnetization dynamics and ferromagnetic resonance.} \textbf{a.} The magnetization \textbf{\textit{M}} (red arrow) precesses about the effective field $\textbf{\textit{H}}_\mathrm{eff}$ (yellow arrow). The field-like torque $\textbf{\textit{M}}\times \textbf{\textit{H}}_\textrm{eff}$ and damping-like torque $\textbf{\textit{M}}\times \frac{d\textbf{\textit{M}}}{dt}$ are shown in blue and purple arrows, respectively. \textbf{b.} Coordinate system used to describe the experimental configuration. The orientation of the applied DC magnetic field ($\Vec{\textbf{\textit{H}}}$) is denoted by ($\theta_H, \phi_H$), and the resulting equilibrium orientation of the magnetization ($\Vec{\textbf{\textit{M}}}$) is given by ($\theta, \phi$) in spherical coordinates. \textbf{c.} Imaginary part of the magnetic susceptibility ($\chi_2$, orange) as a function of magnetic field ($H$) and corresponding magnetic field derivative ($d\chi_{2}/dH$, blue). $H_\mathrm{res}$ is the resonance field and $\Delta H$ is the linewidth.}
\end{figure*}

Magnetization dynamics cover a wide range of phenomena such as magnetization switching, domain wall motion, and the emergence of magnetic textures (vortices, skyrmions, and merons). Here, we consider the precessional magnetization dynamics shown in Fig. \ref{LLG}a, in which the motion of magnetic moment around its equilibrium position is described by the phenomenological Landau-Lifshitz-Gilbert (LLG) equation \cite{gilbert1955lagrangian, lakshmanan2011fascinating}:
\begin{eqnarray}
\frac{d\textbf{\textit{M}}}{dt} = -\gamma \textbf{\textit{M}}\times \textbf{\textit{H}}_\textrm{eff} + \alpha \textbf{\textit{M}}\times\frac{d\textbf{\textit{M}}}{dt}
\label{eq:one}
\end{eqnarray}

\noindent where \textbf{\textit{M}} is the magnetic moment, $\textbf{\textit{H}}_\textrm{eff}$ is the effective field including the external applied DC field, demagnetization field, exchange field, and anisotropy field. The symbol $\gamma=g \mu_{B}/\hbar$ represents the gyromagnetic ratio, and $\alpha$ is the Gilbert damping parameter. For a free electron $g = 2.0023$, one has $\gamma = 1.7588 \times 10^7$ Hz/Oe. The $\textbf{\textit{M}}\times \textbf{\textit{H}}_\textrm{eff}$ term describes the field-like torque that drives the precessional motion of magnetization, while the $M\times \frac{d\textbf{\textit{M}}}{dt}$ term describes the damping-like torque, which damps the precession and aligns the spin towards the effective field.

Magnetic resonance in ferromagnetic materials at microwave frequencies is similar to nuclear and electron spin resonance. The total magnetic moment precesses around the direction of the effective field ($\textbf{\textit{H}}_\textrm{eff}$) at the Larmor frequency. The energy of a small transverse microwave field is absorbed when the frequency of the microwave field coincides with the precession frequency. In this case, the precession angle is maximized, and the material will be able to absorb the maximal microwave power. The FMR resonance frequency is dependent on the external magnetic field, magnetic anisotropy, and temperature. This is equivalent to an additional torque exerted on the electron spin, known as the anti-damping-like torque, which acts in the opposite direction of the damping-like torque. 

The details of FMR theory can be found in review articles ~\cite{farle1998ferromagnetic, maksymov2015broadband, schmool2021ferromagnetic}. Here, the principle of FMR is presented on an intuitive level and the discussion is focused on 2D magnets. Figure \ref{LLG}b shows the typical coordinate system used in FMR experiments, in which ($\theta, \phi$) and ($\theta_H, \phi_H$) are the angles for magnetization ($\vec{\textbf{\textit{M}}}$) and applied magnetic field ($\vec{\textbf{\textit{H}}}$) vector in the spherical coordinates, respectively. As shown in Fig. \ref{LLG}c, the power absorption spectrum is proportional to $\chi_2$ while the field-modulated FMR spectrum is proportional to $d\chi_{2}/dH$, where $\chi_2$ is the imaginary part of the high-frequency magnetic susceptibility. The field dispersive lineshape ~\cite{harder2011analysis} is expressed as 
\begin{eqnarray}
\frac{d\chi_2}{dH} = A_\textrm{sym}\frac{\Delta H^2}{(H-H_\textrm{res})^2 + \Delta H^2} + A_\textrm{asy} \frac{\Delta H^2(H-\Delta H)}{(H-H_\textrm{res})^2 + \Delta H^2}
\label{eq:two}
\end{eqnarray}

\noindent where $H_\mathrm{res}$ is the resonance field, $\Delta H$ is the linewidth, and $A_\mathrm{sym}$ and $A_\mathrm{asy}$ are symmetric and asymmetric Lorentzian components, respectively. 

\subsection{Prototypical FMR techniques}

\begin{figure*}   
\centering
\includegraphics[width=0.8\textwidth]{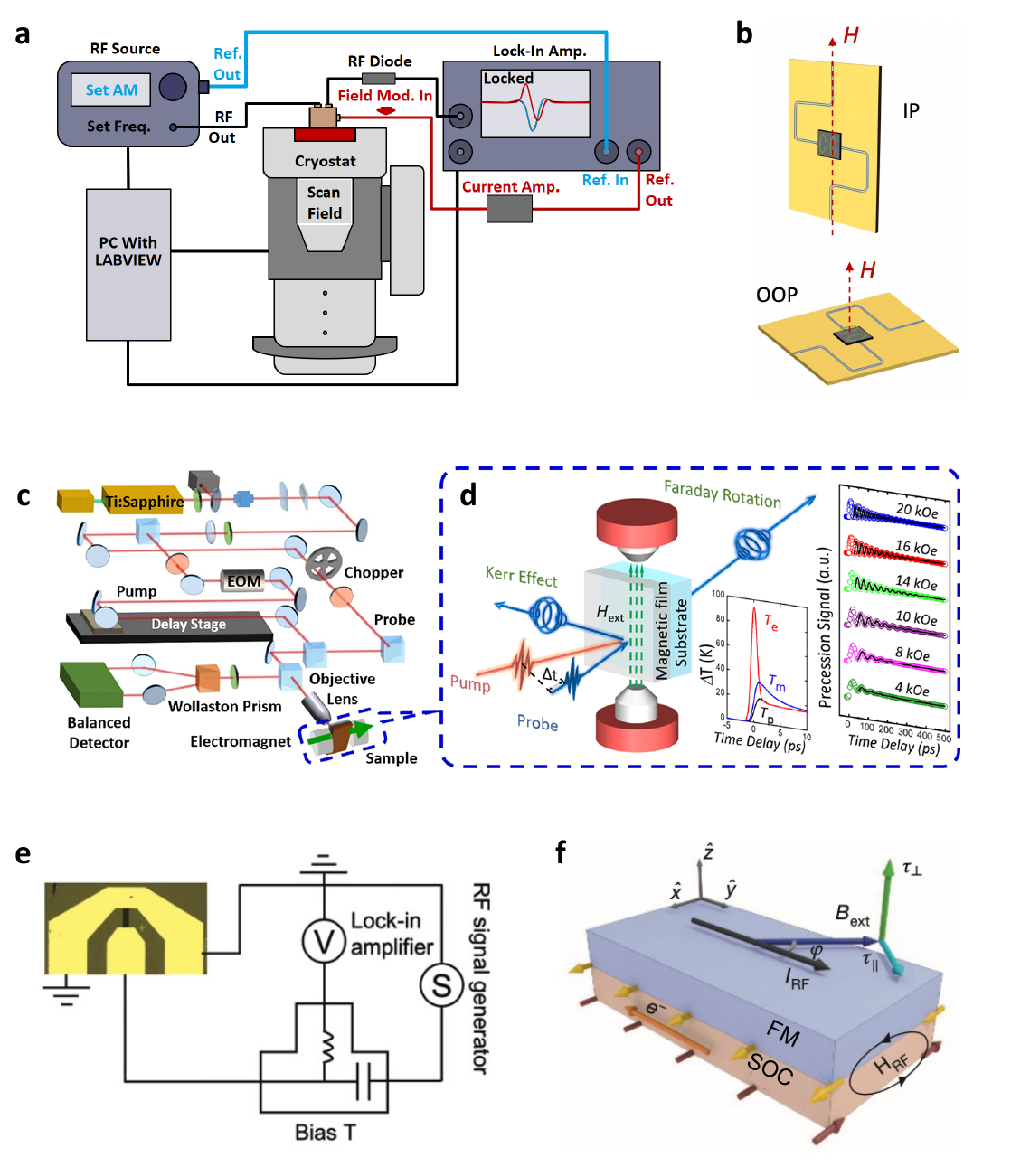}
\caption{\label{Setup} \textbf{Prototypical FMR spectroscopy setups.} \textbf{a.} Main components of broadband FMR include an RF source, a cryostat, an RF diode, and a lock-in amplifier. A controllable DC magnetic field is provided by a physical properties measurement system. The amplitude- and field-modulated spectra are shown in blue and red, respectively. \textbf{b.} Schematic of a co-planar waveguide with a magnetic field applied in the in-plane (IP) and out-of-plane (OOP) geometry, respectively. \textbf{c.} Layout of an example optical FMR setup. A mode-locked Ti:sapphire laser produces a train of pulses with a duration of $\sim$100 fs and a center wavelength of $\sim$800 nm. A polarizing beam splitter separates the laser into pump and probe beams with orthogonal polarizations. The pump beam is modulated as a sinusoidal wave by an electro-optical modulator (EOM). A mechanical delay stage varies the optical path of the pump beam, producing a time separation between the pump excitation and probe sensing. The probe beam is split into two paths of orthogonal polarizations by a Wollaston prism and the changes in the relative intensities of these two probe paths are detected by a balanced detector, which is related to the variations in the polarization states. \textbf{d.} Mechanism of spin dynamics with optical excitation. Both the Faraday rotation (transmission) and the Kerr rotation (reflection) are illustrated, depending on the measurement configuration. Insets show the phenomenological three-temperature model that considers energy exchange between electrons ($T_e$), phonons ($T_p$), and magnons ($T_m$), and the typical pump-probe signal as a function of delay time acquired under varying magnetic fields. \textbf{e.} Optical image of the sample geometry including contact pads, with the circuit used for ST-FMR measurements. \textbf{f.} Schematic diagram of the SOC/FM bilayer structure and coordinate system. The yellow and red arrows denote spin moment directions. $I_\mathrm{RF}$ and $H_\mathrm{RF}$ represent the applied radio frequency current and the corresponding Oersted field. An external magnetic field $B_\mathrm{ext}$ in the film plane at a $\phi$ angle with respect to the current direction. Panels \textbf{c} and \textbf{d} are adapted from Ref. ~\cite{huang2020materials}. Panel \textbf{e} is adapted from Ref. ~\cite{chen2021proximity}. Panel \textbf{f} is adapted from Ref. ~\cite{mellnik2014spin}.}
\end{figure*}

In this review, we focus on three types of FMR techniques: broadband FMR spectroscopy, optical FMR, and spin-torque FMR.

FMR can be used as a spectroscopy method to gain insight into the magnetization dynamics in magnetic materials. A broadband FMR spectroscopy setup is shown in Fig. \ref{Setup}a, which typically is comprised of the following equipment: a controllable DC magnetic field source, such as a physical properties measurement system (PPMS), a radio-frequency (RF) source to supply the required microwave current, a co-planar waveguide (CPW) to deliver microwaves of different frequencies to the sample, an RF diode to convert the returned microwave current into a detectable voltage. Figure \ref{Setup}b shows the typical measurements performed in the in-plane (IP) and out-of-plane (OOP) configuration when the magnetic field is applied along the vertical direction in a PPMS setup. Due to the high-frequency nature, it is difficult to detect RF signals with an adequate signal-to-noise ratio. Therefore, a low-frequency envelope is added to the RF signals, which can be measured using a lock-in amplifier. The low-frequency envelope can be added either through the RF source itself if it supports amplitude modulation (AM), which is highlighted in blue in Fig. \ref{Setup}a. It can also be added externally by supplying an AC signal to a set of Helmholtz coils, which can then modulate the RF current passing through it. This field-modulation option is highlighted in red in Fig. \ref{Setup}a.

Ferromagnetic resonance can also be detected using a set of established optical techniques including magneto-optical Faraday effect, magneto-optical Kerr effect, and transient reflectivity. Optical detection of FMR is typically performed by pump-probe experiments \cite{rasing_rmp2010, hashimoto2017all, mushenok_apl2017, li2019optical,deb_prap2019}, which usually consists of ultrafast lasers, well-aligned opto-mechanical delay-line, high-level vibration control, and balanced detection (see Fig. \ref{Setup}c) ~\cite{zhu2017ultrafast, huang2020materials}. As shown in the insets of Fig. \ref{Setup}d, the changes in the sample's magnetization can be correlated to the Faraday/Kerr rotation angle via the magneto-optical effect, which is dominated by the thermalization processes between different energy carriers, namely, electrons, phonons, and magnons. The Faraday/Kerr signals contain damped oscillating fringes resulting from spin precession, which allows for the analysis of the spin wave modes and damping constant of magnetic materials. Beyond the thermalization picture, in Section 4, we will also discuss the ultrafast magnetization dynamics in vdW magnetic systems through non-thermal effects such as interlayer charge transfer, orbital transition, and electronic structure change.

Spin-torque ferromagnetic resonance (ST-FMR) uses current-induced spin-torques to induce spin precession ~\cite{LQLiu2011spin}. It usually requires the patterning of a spin-orbit-coupling/ferromagnet (SOC/FM) bilayer sample (see Fig. \ref{Setup}f) to strips of micrometer scale. Figure \ref{Setup}a shows the sample configuration with coplanar waveguide patterns for microwave input and signal detection ~\cite{mellnik2014spin}. As shown in Fig. \ref{Setup}b, the microwave-induced an oscillating transverse spin current in the SOC layer. This spin current is injected into the magnetic layer exciting spin precession and generating an RF anisotropic magnetoresistance. This magnetoresistance, rectified with the microwave current, generated a DC voltage. The ST-FMR spectrum  can be obtained by sweeping external in-plane magnetic fields through the resonance condition. The ST-FMR signal ($V_\mathrm{mix}$) can be fitted to the field-derivative lineshape in Eq. (2). The two vector components of the current-induced torque, along the in-plane $\tau_{\parallel} \propto m\times(m\times\sigma) $ and out-of-plane $\tau_{\perp} \propto m\times\sigma$ directions (see Fig. \ref{STFMR}b), are respectively obtained from the amplitudes of the symmetric ($A_\mathrm{sym}$) and anti-symmetric ($A_\mathrm{asy}$) components of the resonance lineshape. As a result, the spin Hall angle can be evaluated from the ratio $A_\mathrm{sym}$/$A_\mathrm{asy}$:

\begin{eqnarray}
\theta_\textrm{SH} = \frac{A_\textrm{sym}}{A_\textrm{asy}}(\frac{e\mu_\textrm{0}M_\textrm{s}t_\mathrm{FM}t_\mathrm{SOC}}{\hbar})[1+\frac{4\pi M_\textrm{eff}}{H_\textrm{ext}}]^{1/2}
\label{eq:SHE}
\end{eqnarray}

\noindent where $e$ is the electron charge, $\mu_\textrm{0}$ is the permeability of vacuum, $M_\textrm{s}$ is the saturation magnetization, $\hbar$ is the reduced Planck's constant, and $t_\mathrm{FM}$ and $t_\mathrm{SOC}$ is the thickness of the magnet and SOC layer, respectively.

It is worth noting that the formalism above can be principally applied to the antiferromagnetic states. For example, in a  simple collinear antiferromagnetic configuration, one can consider the precession of two sublattices having opposite magnetic moments. Within each sublattice, each spin feels identical anisotropy and exchange interaction. Antiferromagnets generally have quite high magnetic resonance frequencies, known as antiferromagnetic resonance, whose frequency can reach up to the sub-THz and THz regimes. The THz magnetization dynamics can be studied using ultrafast magneto-optical experiments. For 2D magnets with layered antiferromagnetic states, the weak interlayer coupling gives rise to a resonance frequency within the typical microwave range, which can be investigated using the broadband FMR technique.

%%%%%%%%%%%%%%%%%%%%%%%%%%%%%%%%%%%%%%%%%%%%%%%%%%%%%%%%%%%%%%%%%%%

\section{Broadband FMR investigation of key parameters of magnetic properties}

\subsection{Magnetic anisotropy}

The resonance frequency can be calculated by considering the classical vector of the macroscopic magnetization (\textbf{\textit{M}}) and the appropriate free energy density ($F$) in the spherical coordinates defined in Fig. \ref{LLG}b ~\cite{artman1957ferromagnetic, baselgia1988derivation}~\cite{Suhl1955Equation, phiilipsSmith}
\begin{eqnarray}
\omega_{res} = \frac{\gamma}{Msin\theta}[\frac{\partial^2 F}{\partial{\theta}^2}\frac{\partial^2 F}{\partial{\phi}^2}-(\frac{\partial^2 F}{\partial\theta\partial\phi})^2]^{1/2}
\label{eq:three}
\end{eqnarray}

\noindent where $F$ in an external magnetic field can be evaluated as the sum of exchange interaction, magnetocrystalline anisotropy, magnetoelastic interaction, demagnetization, Zeeman energy, etc.

\begin{table*}[ht]
\textbf{\caption{The anisotropic part of the free energy density ($F$) for different crystal structures of 2D vdW magnets.}}
\vspace{5pt}
\centering
\renewcommand{\arraystretch}{1.7}
\begin{tabular}{|p{2cm}||p{3cm}|p{3.5cm}|p{8.5cm}|}
\hline
\textbf{Crystal class}  & \textbf{Typical example}  & \textbf{2D magnets}  &   \textbf{Magnetic anisotropy energy}      \\
\hline
Hexagonal & Co & MX$_3$ (M = Cr, V, Ni; X = Cl, Br, I), NPS$_3$ or NPSe$_3$ (N = Mn, Fe, Ni), Cr$_2$Z$_2$Te$_6$ (Z = Ge, Si). &   $K_2sin^2\theta+K_4sin^4\theta+K_{6\perp}sin^6\theta+K_{6\parallel}sin^6\theta cos^6\phi+\cdots$  \\

Trigonal & $\alpha$-\ce{Fe2O3} & \ce{Fe3GeTe2}, \ce{MnBi2Te4}, 2H-\ce{VSe2}, 1T-\ce{VSe2}.   &  $K_2sin^2\theta+K_{41}sin^4\theta+K_{42}sin^3\theta cos\theta cos^3\phi+K_{61}sin^6\theta+K_{62}sin^6\theta cos^6\phi+K_{63}sin^3\theta cos^3\theta cos^3\phi$ \\

Tetragonal  &  Ni,Co,Fe/Cu(001) & FeTe  &  $K_2sin^2\theta+K_{4\perp}sin^4\theta+K_{4\parallel}sin^4\theta cos^4\phi+\cdots$  \\

Orthorhombic & \ce{Fe3O4} & \ce{VOX2} (X= Cl, Br, I)  & $sin^2\theta(K_1cos^2\phi+K_2sin^2\phi)+sin^4\theta(K_3cos^2\theta+K_4sin^2\phi cos^2\phi+K_5sin^4\phi)+\cdots$  \\

 \hline

\end{tabular}
\end{table*}

In general, the magnet moments in 2D vdW magnets originate from the spin and orbital angular momenta of the 3$d$ or 4$f$ electrons in the transition metal ions, which typically interact with the crystal fields associated with crystalline symmetries. Therefore, most of the currently known 2D vdW magnets can be classified into hexagonal, trigonal, tetragonal, and orthorhombic lattices. The anisotropic part of the free energy density ($F$) for these crystal structures is summarized in \textbf{TABLE I}.

Even though the presence of magnetic anisotropy is not a requisite for the appearance of magnetism in strictly two dimensions as recently demonstrated \cite{Jenkins22}, for device implementation and spintronic applications the characterization of a sizeable magnetic anisotropy becomes a key task. FMR spectroscopy is a valuable tool to accomplish this aim because the shift in the resonance field of a few Oersted's can be easily detected, which corresponds to an energy resolution in the determination of magnetic anisotropy on the order of 0.1 $\mu$eV ~\cite{farle1998ferromagnetic}. 

\begin{figure*}[ht!]
\centering
\includegraphics[width=0.8\textwidth]{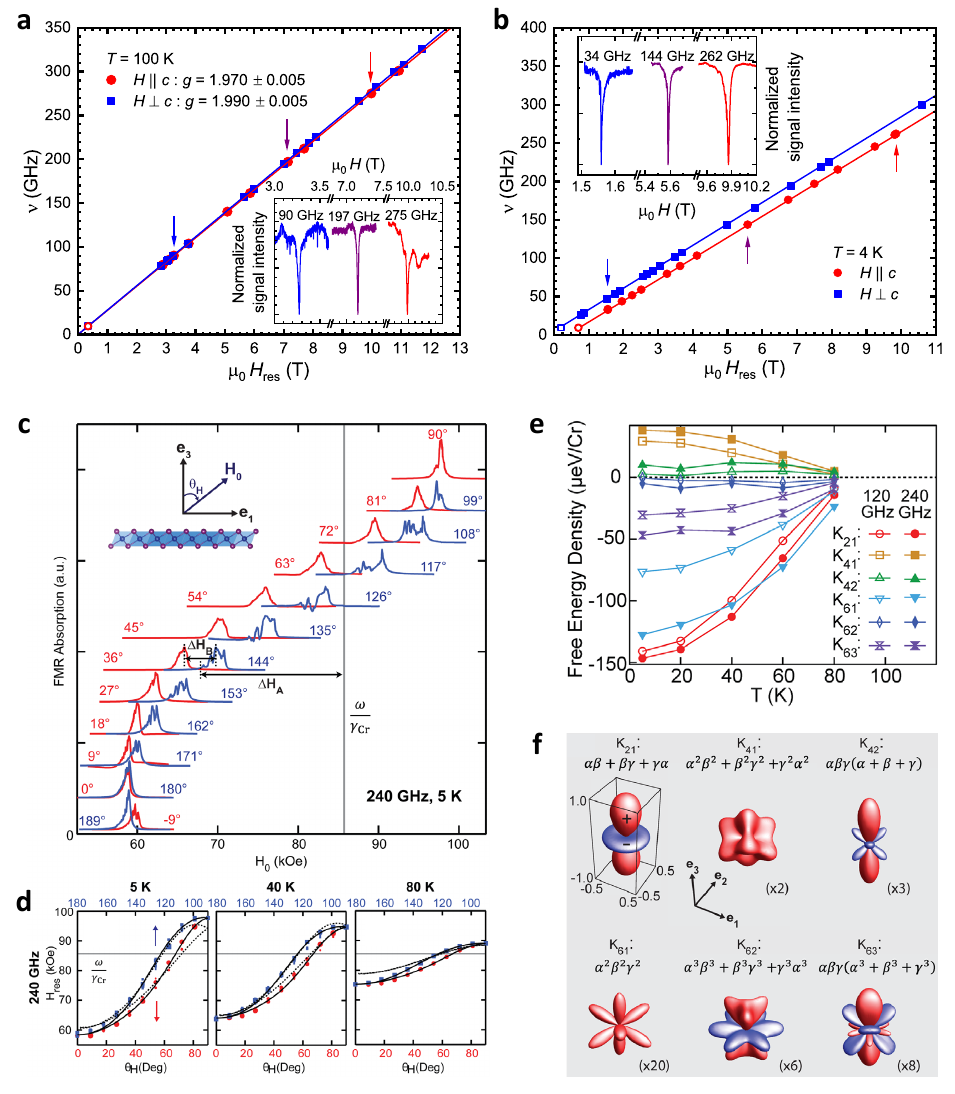}
\caption{\label{Anisotropy} Frequency dependence of the resonance field $H_{res}$ at \textbf{a.} 100 K and \textbf{b.} 4 K, respectively. Measurements were carried out with the external magnetic field applied parallel (red circles) and perpendicular (blue squares) to the crystallographic $c$ axis, respectively.   \textbf{c.} Evolution of the FMR spectra as $\theta_H$ is varied, measured at 240 GHz and 5 K. Each spectrum is offset and scaled moderately for clarity. The same offset is applied for $\theta_H$ and $180^{\circ}-\theta_H$. $\Delta H_A$ and $\Delta H_B$ are two anisotropy features in $H_{res}$. $\omega /\gamma_\mathrm{Cr}$ denotes the corresponding $H_{res}$ for a free Cr$^{3+}$ ion spin. \textbf{d.} $H_{res}$ vs $\theta_H$ for various temperatures. The solid black lines are fits to the free energy function [Eq. (6)]. \textbf{e.} Temperature dependence of the coefficients $K_{pq}$ associated with the basic anisotropy structures. \textbf{f.} Basic anisotropy structure in terms of the cosines $\alpha$, $\beta$, $\gamma$ (the projections of the magnetization onto the x, y, z directions). The sizes are rescaled relative to that for $\alpha\beta+\beta\gamma+\gamma\alpha$ with the indicated magnifications. Red (blue) denotes positive (negative) values. Panels \textbf{a-b} are adapted from Ref. ~\cite{PhysRevMaterials.4.064406}. Panels \textbf{c-f} are adapted from Ref. ~\cite{lee2020fundamental}.}
\end{figure*}

Chromium trihalides (CrX$_3$, X = Cl, Br, I) represent a prototypical vdW magnet, in which the Cr$^{3+}$ cations (3$d^3$, S = 3/2) form a honeycomb structure with the edge-sharing octahedral coordination formed by six halogen anions ~\cite{acharya2021electronic, zhang2022magnetic}. For CrCl$_3$, below the magnetic transition temperature ($T_c$ = 17 K), spins in the honeycomb layers are coupled ferromagnetically and oriented within the \textit{ab} plane, while spins in neighboring layers are coupled by a weaker antiferromagnetic interaction ~\cite{cable1961neutron, mcguire2017magnetic, cai2019atomically}. To determine the details of the magnetic anisotropy in CrCl$_3$, frequency-dependent FMR investigations were conducted at 100 and 4 K ~\cite{PhysRevMaterials.4.064406}. The resulting frequency-field diagrams are shown in the main panels of Fig. \ref{Anisotropy}a and \ref{Anisotropy}b and exemplary spectra are presented in their insets. At 100 K a linear frequency-field dependence of resonance $\nu(H_{res})$ is observed for both orientations of the external magnetic field. Such behavior is expected in the paramagnetic regime of CrCl$_3$ and can be well described by the standard resonance condition of a paramagnet
\begin{eqnarray}
\nu = g\mu_B\mu_0 H_{res}/h 
\label{eq:four}
\end{eqnarray}

\noindent where $g$, $\mu_B$, $\mu_0$ and $h$ denote the $g$-factor, Bohr’s magneton, vacuum permeability, and Planck’s constant, respectively. Fits to the data according to Eq. (4) yielded a $g$-factor with merely small deviation from the free-electron $g$-factor of 2 as well as very slight anisotropy. The result is consistent with the expected value for Cr$^{3+}$ ions and indicates that the magnetism in CrCl$_3$ is largely dominated by the spin degrees of freedom while the orbital angular momentum is completely quenched. Consequently, the spin-orbit coupling and intrinsic magnetocrystalline anisotropies are very weak (or even negligible) in CrCl$_3$, as it was also suggested in the literature ~\cite{mcguire2017magnetic, bastien2019spin}.

In contrast, the data collected at 4 K (see Fig. \ref{Anisotropy}b), i.e., in the magnetically ordered state, shows a clear anisotropy regarding the two magnetic-field orientations. For $H \perp c$, the resonance positions shift towards smaller magnetic fields for all measured frequencies, while for $H \parallel c$, the resonance positions shift to higher magnetic fields. Qualitatively, such behavior reflects a ferromagnetically ordered system with an easy-plane anisotropy. Quantitatively, the sources of magnetic anisotropy can be evaluated using the free energy density below
\begin{eqnarray}
% \begin{split}
\begin{aligned}
F = & -\mu_0 H\cdot M - K_U cos^2(\theta) \\ & +\frac{1}{2}\mu_0 M^2[N_x sin^2(\theta)cos^2(\phi) \\& +N_y sin^2(\theta)sin^2(\phi) + N_z cos^2(\phi)] \label{eq:five}
% \end{split}
\end{aligned}
\end{eqnarray}

\noindent where the first term is the Zeeman-energy density describing the coupling between the magnetization vector and the external magnetic field, the second term represents the uniaxial magnetocrystalline anisotropy whose strength is parameterized by the energy density $K_U$, and the third term is the shape anisotropy energy density which is characterized by the demagnetization factors $N_x$, $N_y$, and $N_z$. An excellent agreement between this model and the measured resonance positions can be accomplished by setting $K_U$ to zero, indicating the magnetic anisotropy is solely due to the shape anisotropy caused by long-range dipole-dipole interactions, whereas the intrinsic magnetocrystalline anisotropy can be neglected. Similar magnetic anisotropy investigation were also performed in the vdW magnet Cr$_2$Ge$_2$Te$_6$~\cite{khan2019spin}.

Comparing with CrCl$_3$, it was concluded that the magnetic anisotropy in CrI$_3$ arises from a dominant uniaxial or single-ion anisotropy ~\cite{dillon1965magnetization, lado2017origin}. However, theoretical proposals and experimental measurements have highlighted the important contributions from high-order exchange interactions (e.g., biquadratic)\cite{Wahab2020,Kartsev2020}, the off-diagonal term ($\Gamma$) and Kitaev interaction ($K$) in the Heisenberg-Kitaev ($J-K-\Gamma$) Hamiltonian ~\cite{xu2018interplay, joshi2018topological, deb2019topological, chen2020magnetic}. The structure of the magnetic anisotropy in CrI$_3$ can be obtained from angle-dependent FMR by measuring the change of the resonance field as the direction of the external field ($H_0$) is varied. Figure \ref{Anisotropy}c shows a representative example of the FMR spectra for different $\theta_H$ at 240 GHz and 5 K. Two distinct features are crucial to analyzing the anisotropy: $\Delta H_\mathrm{A}$ is the shift in $H_\mathrm{res}$ from the free ion contribution $\omega /\gamma_\mathrm{Cr}$, where $\gamma_\mathrm{Cr}$ is the gyromagnetic ratio of Cr$^{3+}$, and $\Delta H_\mathrm{B}$ is the difference in $H_\mathrm{res}$ between $\theta_H$ and $180^{\circ}-\theta_H$. Figure \ref{Anisotropy}d shows the resonance field $H_\mathrm{res}(\theta_H, \omega, T)$ at varying temperatures. A free energy functional is constructed below to fit the data as shown in the black solid curves in Fig. \ref{Anisotropy}d.
% \begin{equation}
\begin{eqnarray}
\label{eq:six}
% \begin{split}
\begin{aligned}
F = & -\mu_0 H\cdot M + 2\pi M^2 cos^2(\theta) + K_{21}(\alpha\beta+\beta\gamma+\gamma\alpha)  \\ & + K_{41}(\alpha^2\beta^2+\beta^2\gamma^2+\gamma^2\alpha^2) + K_{42}\alpha\beta\gamma(\alpha+\beta+\gamma)  \\ & + K_{61}\alpha^2\beta^2\gamma^2 + K_{62}(\alpha^3\beta^3+\beta^3\gamma^3+\gamma^3\alpha^3)  \\ & + K_{63}\alpha\beta\gamma(\alpha^3+\beta^3+\gamma^3)  
% \end{split}
\end{aligned}
\end{eqnarray}
% \end{equation}

\noindent where the first and second term are Zeeman energy and shape anisotropy, respectively. $K_{pq}$ are coefficients associated with the quadrupole interaction. Remarkably, the high spectroscopic precision of FMR enables the evaluation of the $\mu$eV-scale quadrupole interaction constants. The extracted values of the $K_{pq}$ and corresponding basic anisotropy structure are shown in Fig. \ref{Anisotropy}e and \ref{Anisotropy}f, respectively. The results show that Kitaev interaction is the strongest in CrI$_3$, much larger than the Heisenberg exchange, and responsible for opening the gap at the Dirac points in the spin-wave dispersion ~\cite{lee2020fundamental}. Nevertheless, inelastic neutron scattering experiments\cite{Chen2020b,Chen2018,Lebing21} undertaken on bulk CrI$_3$ have pointed out to different conclusions on the magnitude of the Kitaev term influencing the formation of the Dirac gap. That is, the most updated values\cite{Lebing21} resulted in negligible contributions of the Heisenberg-Kitaev Hamiltonian and electron correlation effects to the spin waves and Dirac spin gap in CrI$_3$. The dataset is approximately consistent with a Heisenberg Hamiltonian including anisotropy, exchange interactions and Dzyaloshinskii-Moriya interaction (DMI) with consideration of both the $c$ axis and in-plane DMI. Additional FMR experiments would be needed to fully address this discrepancy. 

\subsection{Anisotropic $g$-factor}

\begin{figure*}
\includegraphics[width=1.0\textwidth]{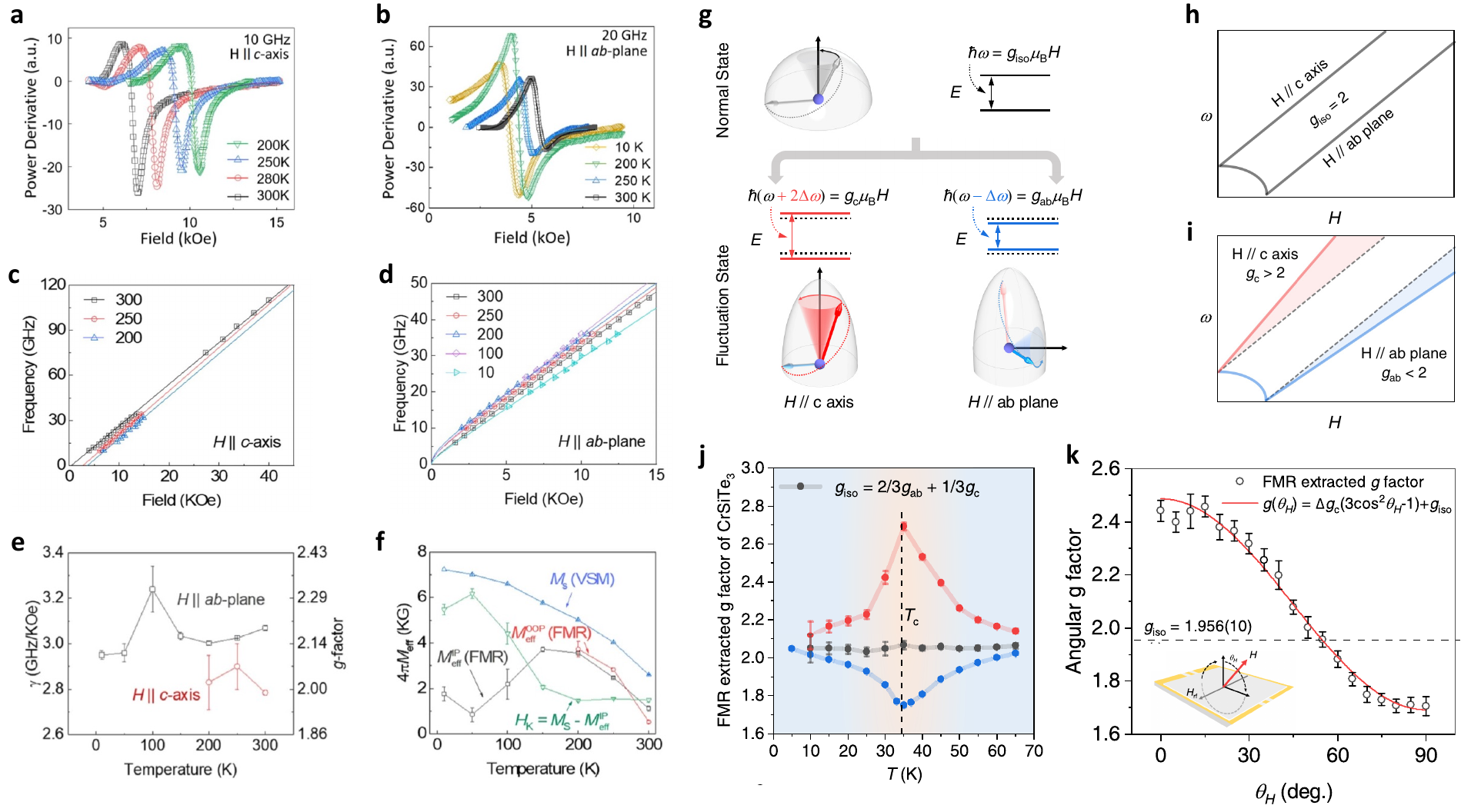}
\caption{\label{g-factor} FMR spectra of \ce{Fe5GeTe2} acquired at varying temperatures for \textbf{a.} $H \parallel c$-axis and \textbf{b.} $H \parallel ab$-plane. The raw data in \textbf{f} and \textbf{g} are fitted to field-derivative lineshape using Eq. (2). \textbf{c.} and \textbf{d}. Frequency vs. resonance field at varying temperatures for $H \parallel c$-axis and $H \parallel ab$-plane, respectively. The data points are fitted to corresponding Kittel equations. \textbf{e}. Temperature dependence of the gyromagnetic ratio $\gamma$ and spectroscopic \textit{g}-factor for the \textit{H}$\parallel$\textit{c} (red) and \textit{H}$\parallel$\textit{ab} (black) cases, respectively. \textbf{f}. Temperature dependence of saturation magnetization 4$\pi$M$_\textrm{s}$ and effective magnetization 4$\pi$M$_\textrm{eff}$ from VSM and FMR measurements, respectively. \textbf{g.} Illustration of the anisotropic $g$-factor, the shifts of energy diagrams, as well as the magnetic torque dynamics around an effective magnetic field in normal and fluctuation states. FMR spectra above the saturation field will give \textbf{h} parallel straight lines along easy ($H \parallel c$ axis) and hard ($H \parallel ab$ plane) direction for isotropic $g_\mathrm{iso} = 2$, in contrast to \textbf{i} the unparallel behavior for anisotropic $g_\mathrm{c} > 2$ and $g_\mathrm{ab} < 2$. \textbf{j.} Comparison of the temperature-dependent $g$-factor extracted from FMR data above saturation field. Calculation of isotropic component $g_\mathrm{iso}$ = 2/3$g_\mathrm{ab}$ + 1/3$g_\mathrm{c}$ is basically in line with the spin-only value $g$ = 2. \textbf{k.} Angular dependence of the extracted $g$-factor. The red line is fitted by $g(\theta_H) = \Delta g(3\mathrm{cos}^2 \theta_{H} - 1)+g_\mathrm{iso}$. Inset shows the geometrical configuration of 3D vector FMR experiment. Panels \textbf{a-f} are adapted from Ref. ~\cite{alahmed2021magnetism}. Panels \textbf{g-k} are adapted from Ref. ~\cite{li2022anomalous}.}
\end{figure*}

Understanding the mechanism of ferromagnetism in metals has been a longstanding nontrivial question in a wide variety of condensed-matter systems ~\cite{korenman1977local, capellmann1979theory, moriya1979recent, moriya1991theory}. Along with the rapidly growing interest in searching for novel vdW magnets, one particular issue has been clarifying whether an orbital with magnetic moment should be considered in a localized or itinerant picture because long-range magnetism starting from localized isotropic Heisenberg spins is not allowed in two dimensions. Fe$_\mathrm{n}$GeTe$_2$ (n = 3, 4, 5) represent a class of metallic vdW ferromagnets with high Curie temperature ~\cite{deng2018gate, seo2020nearly, zhang2020itinerant, mondal2021critical}. Among them, Fe$_5$GeTe$_2$ has a complex atomic structure with multiple nonequivalent iron sites \cite{zhang2020itinerant, wu2021direct, ershadrad2022unusual}. Recent theoretical calculations have shown that the orbital states associated with the nonequivalent iron sites have distinct impact on the magnetic ground state ~\cite{joe2019first}. In addition, x-ray magnetic circular dichroism studies of elemental selective valence-band electronic and spin states highlight significant ligand states (Ge 4$p$ and Te 5$p$) contribution to the ferromagnetism through hybridization with the Fe 3$d$ orbital ~\cite{yamagami2021itinerant}.

The mixture of orbital momentum and magnetic moment  causes the shift of $g$-factor from the free electron value, which may be different along different crystallographic directions, that is, one has to consider a $g$-tensor. This can be demonstrated by writing the magnetic moment $M = -\mu_B (L+2S) \rightarrow g\mu_B S'$, where $S'$ is an effective spin. The contribution of orbital momentum contained in the $g$-tensor can be obtained by direct measurements of $g$-factor using FMR spectroscopy. In the best cases, $g$-factors of metallic ferromagnets are usually quoted with an error bar of 1\% ~\cite{farle1998ferromagnetic}.

FMR spectroscopy studies of the magnetization dynamics of Fe$_5$GeTe$_2$ are summarized in Fig. \ref{g-factor}a-\ref{g-factor}f ~\cite{alahmed2021magnetism}. The FMR spectra for the $H\parallel c$-axis and $H\parallel ab$-plane at selected values of temperature are shown in Fig. \ref{g-factor}a and \ref{g-factor}b, respectively. The profiles were fitted to the field-derivative lineshape using Eq. (2) to extract the resonance fields. The microwave frequencies ($f$) as a function of extracted resonance fields ($H_\mathrm{res}$) were plotted in Fig. \ref{g-factor}c and \ref{g-factor}d. The gyromagnetic ratio ($\gamma$) and corresponding $g$-factor, along with the effective magnetization $4\pi M_\textit{eff}$ can be obtained by fitting the $f$ vs. $H_\mathrm{res}$ data to the Kittel equations as follows:
\begin{equation}
\begin{split}
  & H \parallel c: \hspace{5mm} f_\mathrm{OOP} = \frac{\gamma}{2\pi} (H_\mathrm{res} -4\pi M_\textrm{eff}) \\
  & H \parallel ab: \hspace{3mm} f_\mathrm{IP} = \frac{\gamma}{2\pi} \sqrt{(H_\mathrm{res} +4\pi M_\textrm{eff}) H_\mathrm{res}}
\end{split}
\end{equation}

\noindent The gyromagnetic ratio and corresponding $g$-factor for both field orientations are shown in Fig. \ref{g-factor}e, revealing an anisotropic $g$-factor that, in both cases, deviates from $g$ = 2. The anisotropic $g$-factor can be interpreted physically: a small orbital moment arising from reduced crystalline symmetry may lock the large isotropic spin moment into its favorable lattice orientation through spin–orbit coupling, giving rise to a sizable magnetic anisotropy. Therefore, it is likely that the orbital moment is closely linked to the magnetocrystalline anisotropy in itinerant ferromagnets.

It is worth noting that an unsaturated magnetization at FMR can also lead to an inaccurate estimation of the gyromagnetic ratio. The FMR-extracted effective magnetization, which includes the saturation magnetization as well as the magnetocrystalline anisotropy field $H_\textrm{k}$ can be comparable with the saturation magnetization extracted from the VSM measurements, as shown in Fig. \ref{g-factor}f. To confirm the $g$-factor anisotropy, FMR measurements at high-frequency and high-magnetic-field are needed. 

Beyond the spin-orbit coupling-induced $g$-factor shift at zero field, the Nagata theory of anisotropic magnets ~\cite{nagata1977short, nagata1977epr} indicates that a correction of $g$-factor shift under high-temperature perturbation approximation and in the presence of finite magnetic field should be considered for the studies of critical phenomena in low-dimensional magnet. Recently, the temperature dependence of $g$-factor in ferromagnetic semiconductor \ce{CrSiTe3} has been investigated using broadband three-dimensional vector-ferromagnetic resonance experiments ~\cite{li2022anomalous}. The rectangular-shaped sample is placed in a coplanar waveguide where an alternating magnetic field ($H_\mathrm{rf}$) is applied. In response to scanning a static magnetic field ($H$) at the right angle, resonance absorption occurs in the case that $\hbar\omega=g\mu_B H$. As shown in Fig. \ref{g-factor}g, in the normal state, the Zeeman splitting ($g_\mathrm{iso} = 2$) is independent of orientation, and the FMR spectra above the saturation field (Fig. \ref{g-factor}h) will give parallel straight lines along the easy ($H \parallel c$-axis) and hard ($H \parallel ab$-plane) direction. In the fluctuation state, Nagata theory indicates the $g$-factor shift along easy axis $\Delta g_\mathrm{c}$ is twice the value along the hard plane $\Delta g_\mathrm{ab}$. Therefore, FMR spectra above the saturation field (Fig. \ref{g-factor}i) will show the upwards- (red) or downwards (blue) shifts of the slopes, which clearly indicate the $g$-factor shift.

As shown in Fig. \ref{g-factor}g, the clear downwards- or upwards shifts of the $g$-factor, with a maximum value at $T_c = 34.15$ K, are observed at varied temperatures. In addition, the cryogenic vector magnet system allows angular dependent measurements of $g$-factor at $T_c$. As shown in Fig. \ref{g-factor}h, the $(3\mathrm{cos}^2 \theta_{H} - 1)$ like angular dependence derived from the Nagata theory fits well with the FMR extracted $g$-factor, where the magic angle with a spin-only $g$-factor is observed at $\theta_H = 54.74^{\circ}$. Here, the anisotropic components are taken into account in the spin Hamiltonian of \ce{CrSiTe3}. In contrast, \ce{CrGeTe3} with a similar crystalline structure but quasi-isotropic Heisenberg spin model exhibits a much smaller $g$-factor shift. Therefore, the shift of $g$-factor near the critical temperature in \ce{CrSiTe3} is essentially attributed to both the enhanced magnetic fluctuations and anisotropic spin interactions.

\subsection{Effect of Magnetic domain structures}

The broadband FMR experiments can be combined with micromagnetic and atomistic simulations to investigate the magnetic properties of vdW magnets associated with domain structures. In complement to the microscopy techniques such as Lorentz transmission electron microscopy and magneto-optical microscopy, FMR results emphasize on the dynamical properties which provide essential information for device applications such as magnetic switching, magnon excitation, and damping mechanism. An exemplary work is an FMR measurement of magnetization dynamics of CrI$_3$ and CrBr$_3$ crystals in a broad frequency range of 1 – 40 GHz and over a wide temperature range of 10 -- 300 K ~\cite{Chunli_Ref.7shen2021multi}.

\begin{figure*}[ht!]
\includegraphics[width=1.0\textwidth]{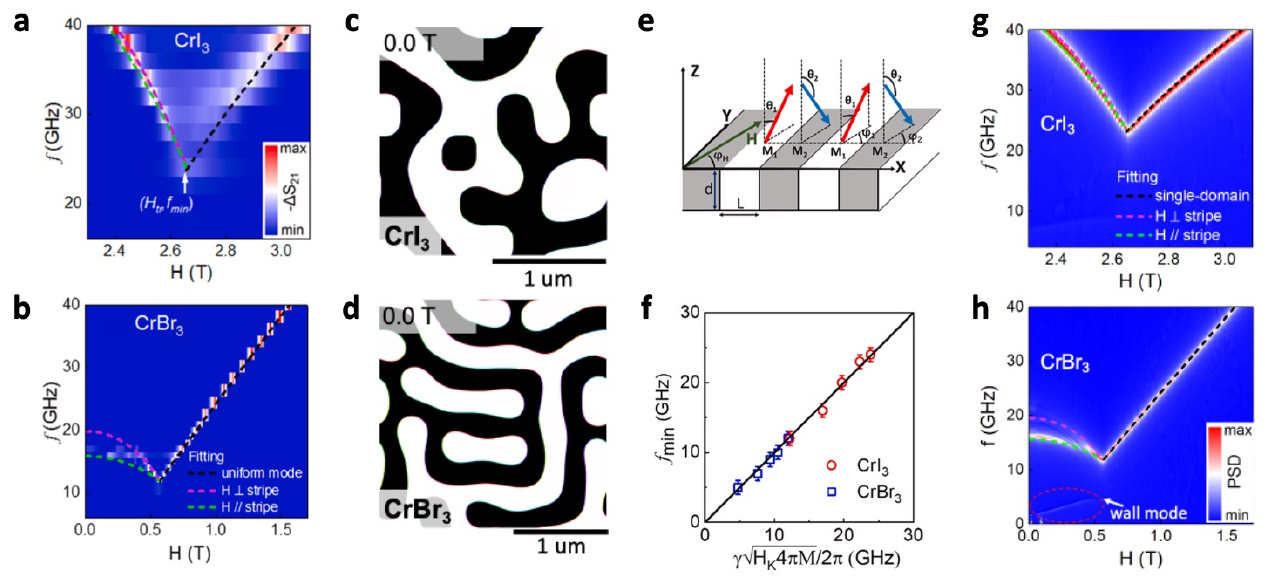}
\caption{\label{Domain} The color map of FMR spectra at 10 K for \textbf{a.} CrI$_3$ and \textbf{b.} CrBr$_3$. The dash lines are fitting curves of the uniform mode (black) and the calculated curves using multi-domain FMR theory (red and green). The simulated domain structures at 10 K and zero field for \textbf{c.} CrI$_3$ and \textbf{d.} CrBr$_3$. \textbf{e.} The illustration of magnetization structures for the stripe domain phase with the in-plane field for the multi-domain FMR calculation. \textbf{f.} The correlation between the experimental values of $f_\mathrm{min}$ and $\gamma/2\pi\sqrt{H_K \cdot 4\pi M}$. The solid line is calculated by Eq. (11).
The color map of simulated FMR spectra utilizing the magnetic parameters for \textbf{g.} CrI$_3$ and \textbf{h.} CrBr$_3$ at 10 K. All panels are adapted from
Ref. \cite{Chunli_Ref.7shen2021multi}.}
\end{figure*}

The color maps in Fig. \ref{Domain}a (CrI$_3$) and \ref{Domain}b (CrBr$_3$) show that the resonance frequency first decreases with $H$ towards a critic field $H_\mathrm{tr}$, then increases with $H$ at higher field, giving rise to a minimum resonance frequency $f_\mathrm{min}$ at $H_\mathrm{tr}$. A quantitative analysis of the FMR features is based on the periodic stripe domain structure which can be reproduced via micromagnetic (see Fig. \ref{Domain}c and \ref{Domain}d) and atomistic simulations\cite{Wahab2020,Wahab21}. The sample consists of two sets of domains with equal volume, and the magnetization in the neighboring domain should have the opposite magnetization component along $z$-direction, thus there is no net $M_z$ under the in-plane magnetic field. The free energy density for the multi-domain structure can be written as
\begin{eqnarray}
% \begin{split}
\begin{aligned}
F = & -\frac{H\cdot M}{2}(sin\theta_1 sin\phi_1+sin\theta_2 sin\phi_2) \\ &+ \frac{K_z}{2}(sin^2\theta_1+sin^2\theta_2) + \frac{4\pi}{2}\frac{M^2}{4}(cos\theta_1+cos\theta_2)^2\\ &+ \frac{N_x}{2}\frac{M^2}{4}(sin\theta_1 cos\phi_1-sin\theta_2 cos\phi_2)^2 \\ &+\frac{N_y}{2}\frac{M^2}{4}(sin\theta_1 sin\phi_1-sin\theta_2 sin\phi_2)^2 \\ &+\frac{N_z}{2}\frac{M^2}{4}(cos\theta_1-cos\theta_2)^2
% \end{split}
\end{aligned}
\label{eq:eight}
\end{eqnarray}

\noindent where $M$ represents the saturation magnetization of material, $\theta_1$ and $\theta_2$ represent the out-of-plane angles of domain magnetization, $\phi_1$ and $\phi_2$ represent the in-plane projection angles of magnetization in each domain, $K_z$ is the perpendicular magnetic anisotropy energy density. $(N_x, N_y, N_z)$ are the demagnetization parameters of each domain with the relation of $N_x + N_y + N_z = 4\pi$. Figure \ref{Domain}e shows that the domain system follows the conditions of $\theta_1+\theta_2=\pi$ and $\phi_1=\phi_2=\phi_H$. The resonance frequency $f$ in the multi-domain state depends on the in-plane field angle $\phi_H$. For $\phi_H=0^{\circ}$ with $H\perp$ strip, the FMR frequency can be described by 
\begin{equation}
    f = \frac{\gamma}{2\pi}\sqrt{H(H-H_K+4\pi M)}
\end{equation}

\noindent and for $\phi_H=90^{\circ}$ with $H\parallel$ strip, the FMR frequency can be described by 
\begin{equation}
    f = \frac{\gamma}{2\pi}\sqrt{H_{K}^2 + 4\pi M \cdot H^2 /H_K - H^2}
\end{equation}

\begin{figure*}[ht]
 \centering
 \includegraphics[width=1.0\textwidth]{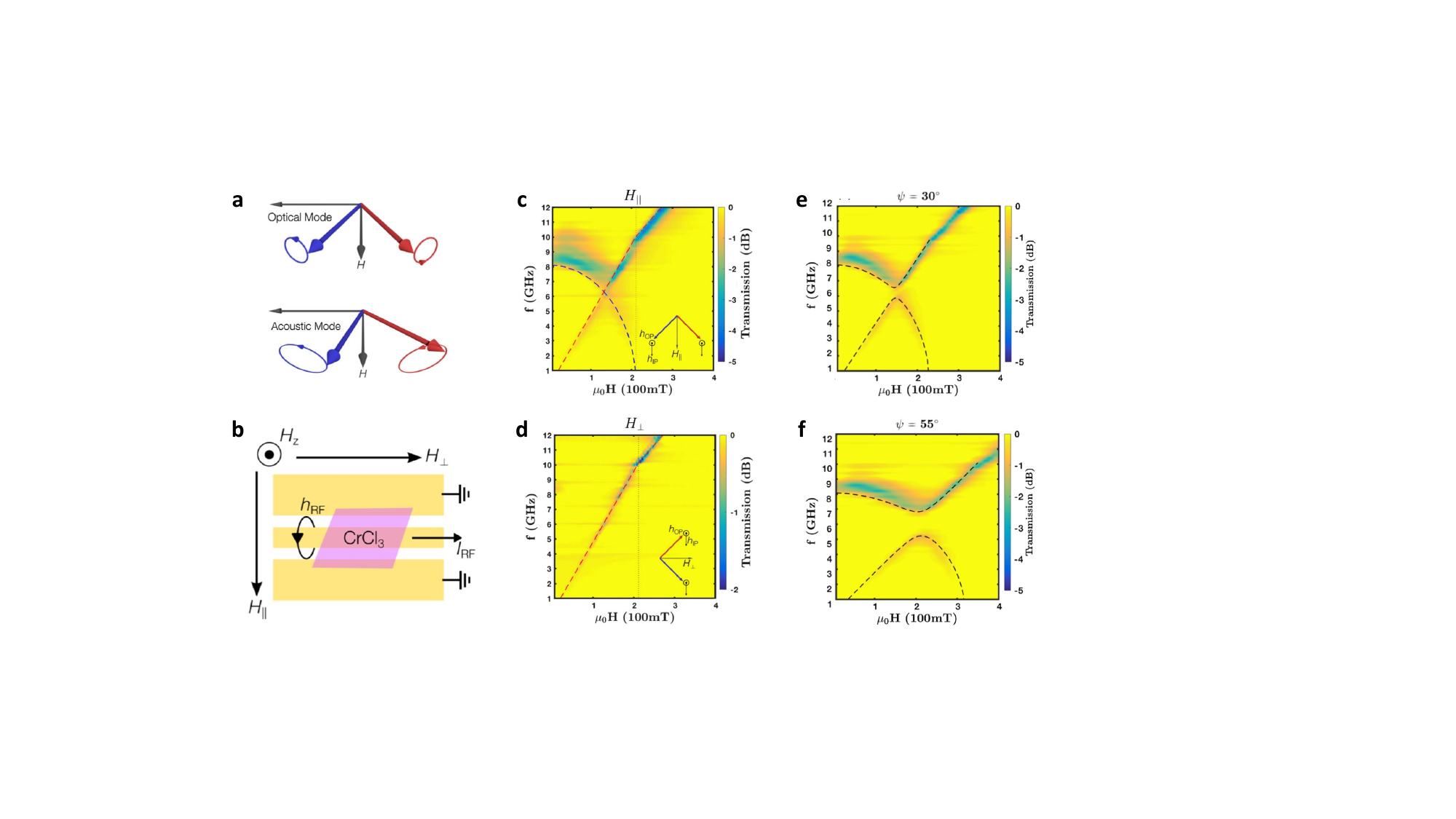}
 \caption{\textbf{a.} Schematic of the precession orbits for the two sublattice magnetizations in the optical mode and the acoustic mode. \textbf{b.} Experimental schematic featuring a coplanar waveguide with a CrCl$_3$ crystal placed over the signal line. $H_{\parallel}$, $H_{\perp}$, and $H_z$ are the components of the applied DC magnetic field. Microwave transmission as a function of frequency and in-plane magnetic field at 1.56 K with the magnetic field applied \textbf{c.} parallel and \textbf{d.} perpendicular to the in-plane RF field. Two modes are observed in the $H_{\parallel}$ configuration: an optical mode that has finite frequency at zero applied field, and an acoustic mode with frequency proportional to the applied field. Only the acoustic mode is observed in the $H_{\perp}$ configuration. Out-of-plane field applied at an angle of \textbf{e.} $\psi = 30^{\circ}$ and \textbf{f.} $\psi = 55^{\circ}$ breaks the symmetry and couples the two modes resulting in tunable gaps. All panels are adapted from Ref. ~\cite{macneill2019gigahertz}.}
 \label{Gap}
\end{figure*}

The relative angle $\phi_H$ between the stripe domain and the applied field could vary between $0^{\circ}$ and $90^{\circ}$, then the corresponding FMR frequency should vary between the values calculated by Eq. (9) and Eq. (10). As a result, the minimum FMR frequency $f_\mathrm{min}$ is obtained when the field $H_{tr}$ equal to $H_K$
\begin{equation}
    f_{min} = \frac{\gamma}{2\pi}\sqrt{H_K \cdot 4\pi M}
\end{equation}

Thus, $f_\mathrm{min}$ can be calculated by Eq. (11) with experimentally determined $\gamma$, $H_K$, and $4\pi M$, as shown in Fig. \ref{Domain}f. Starting from the labyrinth domain structure at 10 K and zero field,  field-dependent evolution of domain structures can be simulated by gradually applying the in-plane field along $x$-axis. The color maps (Fig. \ref{Domain}g and \ref{Domain}h) generated from the simulation agree with the experimental results (Fig. \ref{Domain}a and \ref{Domain}b) very well. Similar color maps of frequency-dependent ferromagnetic resonance spectra and multi-domain structure have also been demonstrated in Cr$_2$Ge$_2$Te$_6$~\cite{khan2019spin}.

%%%%%%%%%%%%%%%%%%%%%%%%%%%%%%%%%%%%%%%%%%%%%%%

\subsection{Magnon-magnon coupling}

CrI$_3$ and CrCl$_3$ have been shown to be layered antiferromagnetic insulators in their few-layer form, in which spins within each layer have a ferromagnetic nearest-neighbor coupling, whereas spins in adjacent layers have a weak antiferromagnetic coupling ~\cite{wang2011electronic, huang2017layer, cai2019atomically, wang2019determining}. Because of the weak interlayer antiferromagnetic coupling in CrCl$_3$ is within the typical microwave range, it allows to excite both acoustic and optical modes (see Fig. \ref{Gap}a) ~\cite{macneill2019gigahertz}. Magnetic resonance measurements were carried out by fixing the excitation frequency and sweeping the applied magnetic field (see experimental geometry in Fig. \ref{Gap}b). Only one resonance is observed when the DC magnetic field is applied perpendicular to the RF field ($H_{\perp}$, see Fig. \ref{Gap}d), but two resonances show up when the DC magnetic field is applied parallel to the RF field ($H_{\parallel}$, see Fig. \ref{Gap}c). The optical mode and acoustic mode are centered around the frequencies $\omega_{\pm}$ with magnetic field dependence as follows:
\begin{eqnarray}
% \begin{split}
\begin{aligned}
&\omega_{+} = \mu_0\gamma\sqrt{2H_E M_s(1-\frac{H^2}{4H_{E}^2})} \\
&\omega_{-} = \mu_0\gamma\sqrt{2H_E(2H_E + M_s)}(\frac{H}{2H_E})
% \end{split}
\end{aligned}
\end{eqnarray}

\noindent where $H_E$ is the interlayer exchange field and $M_s$ is the saturation magnetization. The two modes in Fig. \ref{Gap}a can be fitted to Eq. (12), which are shown by the dashed lines in Fig. \ref{Gap}c with fit parameters of $\mu_0 H_E$ = 105 mT and $\mu_0 M_s$ = 396 mT at $T$ = 1.56 K. Note that the acoustic mode changes its slope at $\mu_0 H \approx$ 200 mT. This occurs because the moments of the two sublattices are aligned with the applied field when $H > 2H_E$. In this case the crystal behaves as a ferromagnet and the acoustic mode transforms into uniform ferromagnetic resonance described by the Kittel formula. 

When the magnetic field is applied in plane, the system is symmetric under two-fold rotation around the applied field direction combined with sublattice exchange. This prevents hybridization between the optical and acoustic modes, leading to a degeneracy where they cross. To break the rotational symmetry, a magnetic field can be applied at a range of angles, $\psi$, from the coplanar waveguide. For $\psi = 30^{\circ}$, a gap opens near the crossing point (see Fig. \ref{Gap}e). Increasing the tilt angle ($\psi = 55^{\circ}$) increases the gap size as shown in Fig. \ref{Gap}f until the mode coupling becomes zero again when $\psi$ approaches $90^{\circ}$. The strong magnon-magnon coupling in CrCl$_3$ with a large tunable gap up to 1.37 GHz may hold the promise of potential applications in magnon-photon, magnon-qubit hybrid quantum systems in a microwave cavity ~\cite{PhysRevLett.113.156401, PhysRevLett.114.227201, tabuchi2015coherent_qubit}.

%%%%%%%%%%%%%%%%%%%%%%%%%%%%%%%%%%%%%%%%%%%%%%%

\section{Optical excitation and detection of magnetization dynamics}

The demand for high-speed operation of spintronic devices has triggered an intense search for ways to control magnetization by ultrafast lasers. In this section, we will discuss laser-induced demagnetization and spin dynamics in 2D magnets that are directly related to the thermalization process. Moreover, we will cover the ultrafast magnetization dynamics through non-thermal effects, which has been demonstrated in a variety of materials ~\cite{kimel2005ultrafast, satoh2010spin, satoh2012directional, satoh2015writing, bossini2016macrospin, stupakiewicz2017ultrafast, nova2017effective} but remains less explored in vdW magnetic systems.

\subsection{Pump-induced magnetization dynamics}

Recently, magnetization dynamics in Cr$_2$Ge$_2$Te$_6$ nanoflakes have been studied by means of a two-color time-resolved Faraday rotation with a picosecond time resolution at cryogenic temperature ~\cite{zhang2020laser}. Below the Curie temperature, a pump beam of femtosecond pulses intensively heats the ferromagnet to produce instantaneous demagnetization and trigger the magnetization precession. By monitoring pump-induced magnetization dynamics in an external magnetic field $H_\mathrm{ext}$, the angular dependence of the precession frequency and the transverse spin relaxation time can be determined.  

\begin{figure*}[ht!]
 \centering
 \includegraphics[width=0.85\textwidth]{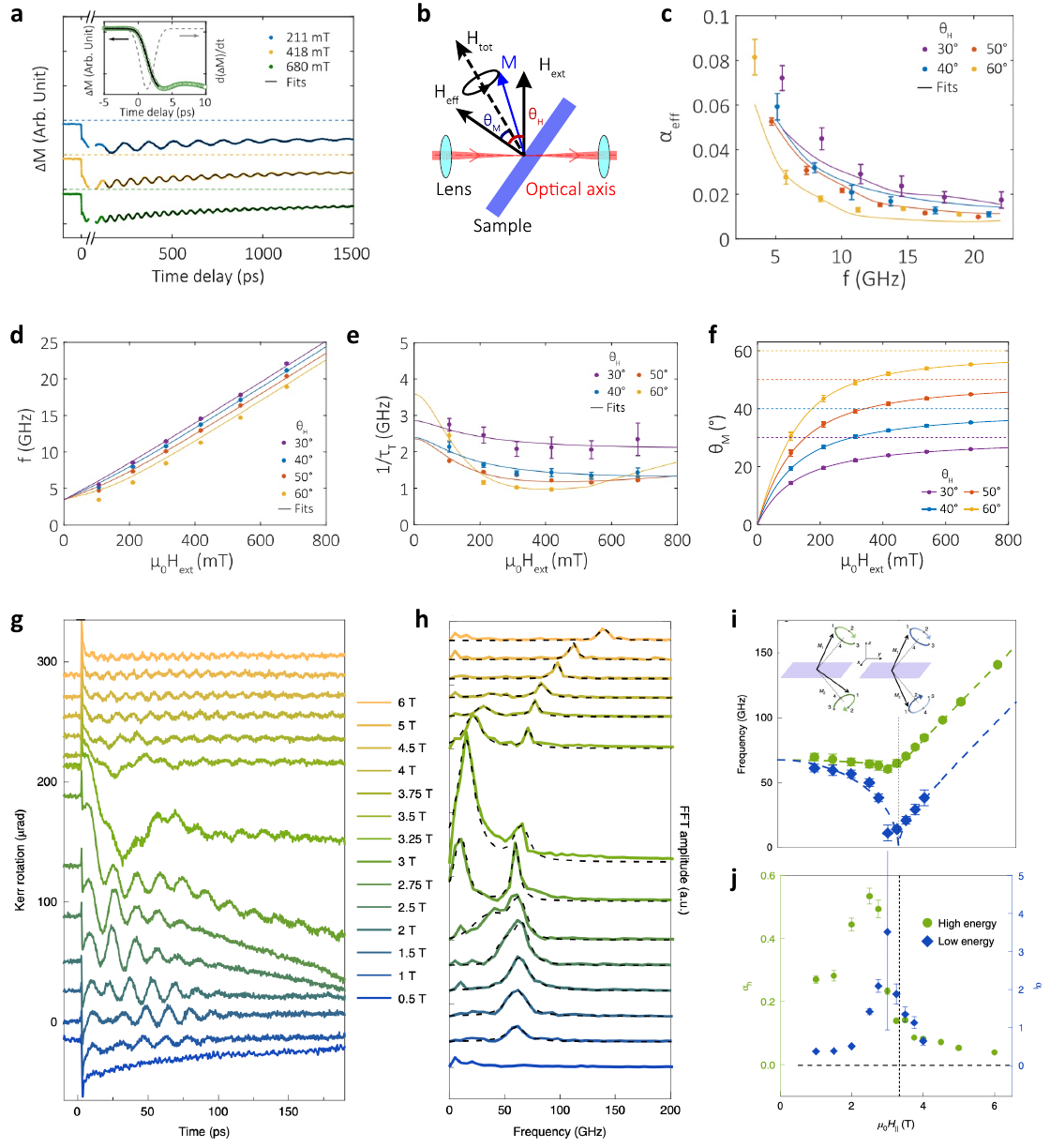}
 \caption{\textbf{a.} Normalized magnetization dynamics due to laser heating in Cr$_2$Ge$_2$Te$_6$ for selected $H_\mathrm{ext}$ (211, 418, and 680 mT) at $\theta_H=50^{\circ}$. Solid lines are fits to Eq. (13) and dashed lines are baselines in the absence of pumping. Inset: Magnetization dynamics (markers) and cumulative Gauss fit (line) in the first 10 ps (left axis). Dashed line (right axis) is the differential curve of the fit, which gives a time resolution of 1.2 $\pm$ 0.01 ps. \textbf{b.} Schematic of the experimental geometry. The sample normal is rotated through the angle $\theta_H$ with respect to $H_\mathrm{ext}$, which is always perpendicular to the optical axis. The angle $\theta_M$ denotes the magnetization that deviates from the easy axis. \textbf{c.} Effective damping $\alpha_\mathrm{eff}$ calculated from $\tau_{T}$, $H_1$, and $H_2$ vs precession frequency $f$ at varied $\theta_H$ (markers). Solid lines are fits to Eq. (14). \textbf{d.} Precession frequency ($f$), \textbf{e.} transverse spin relaxation rate ($\tau_{T}^{-1}$), and \textbf{f.} equilibrium angle ($\theta_M$) as a function of $H_\mathrm{ext}$ at changing $\theta_H$ of $30^{\circ}$, $40^{\circ}$, $50^{\circ}$, and $60^{\circ}$. Markers denote experimental data and solid lines are fitted to the Kittel equation. \textbf{g.} Kerr rotation as a function of pump–probe delay time in bilayer CrI$_3$ under different in-plane magnetic fields and \textbf{h.} corresponding fast Fourier transform (FFT) spectra after the removal of the demagnetization dynamics (exponential decay). The curves are vertically displaced for clarity. \textbf{i.} Frequency and \textbf{j.} damping rates of high energy ($\omega_h$) and low energy ($\omega_l$) modes as a function of in-plane magnetic field. Dashed lines in \textbf{g} are fits to the LLG equation. Inset of \textbf{i} shows the spin precession eigenmodes of $\omega_h$ and $\omega_l$ under an in-plane field (y axis). The vertical dotted lines in \textbf{i} and \textbf{j} indicate the in-plane saturation field from the fits to the LLG equation. Panels \textbf{a-f} are adapted from Ref. ~\cite{zhang2020laser}. Panels \textbf{g-j} are adapted from Ref. ~\cite{zhang2020gate}.}
 \label{PP}
\end{figure*}

Figure \ref{PP}a plots the pump-induced magnetization dynamics up to 1500 ps for $\theta_H=50^{\circ}$ at three external fields $H_\mathrm{ext}$ = 211, 418, and 680 mT. The time resolution of the pump–probe measurements is estimated to be 1.2 $\pm$ 0.01 ps (see inset of Fig. \ref{PP}a). The experimental geometry is schematically described in Fig. \ref{PP}b. In an equilibrium state, the static magnetization remains along the total effective field $H_\mathrm{tot}=H_\mathrm{eff}+H_\mathrm{ext}$. The deviation of magnetization from the easy axis ($\theta_M$) thus reaches $\theta_H$ by increasing $H_\mathrm{ext}$. The ultrafast magnetization dynamics can be qualitatively understood by considering a drastic increase in the spin temperature leading to a reduction of both the magnetization $M$ and the effective magnetic field $H_\mathrm{eff}$. The total magnetic field ($H_\mathrm{tot}$) is modified to tilt the magnetization ($M$) away from its equilibrium direction. After spin cooling of a few picoseconds, $M$ recovers its magnetization saturation value while the electron temperature cools through phonon emission. The change of magnetization as a function of time delay $\Delta M(\Delta t)$ probed by the Faraday rotation can be phenomenally described as 
\begin{eqnarray}
\Delta M(\Delta t) \sim & A_0 e^{-\delta t/\tau_{T}}cos(2\pi f \Delta t + \phi_0) \nonumber \\ &+ A_1 e^{-\delta t/\tau_{p1}} + A_2 e^{-\delta t/\tau_{p2}}
\end{eqnarray}

\noindent Here, the first term describes the spin-precession dynamics in which $A_0$, $\tau_{T}$, $f$, and $\phi_0$ denote the magnetization amplitude, the transverse spin relaxation time, the precession frequency, and the initial phase, respectively. The second and third terms describe the magnetization recovery process, where $A_1$ ($A_2$) and $\tau_{p1}$ ($\tau_{p2}$) are, respectively, the demagnetization magnitude and the characteristic time constant of a short (long) process. By fitting the experimental data to Eq. (13), one can obtain the precession frequency, oscillation magnitudes, and magnetization recovery time. The frequency $f$ and the transverse spin relaxation rate $\tau_{T}^{-1}$ as a function of the external magnetic field are shown in Fig. \ref{PP}d and \ref{PP}e, respectively. All the $f$ vs $H_\mathrm{res}$ can be fitted to a Kittel equation, yielding $\mu_0 H_K = 125$ $\pm$ 8 mT and a $g$-factor of 2.04 $\pm$ 0.03. While $\tau_{T}^{-1}$ at $\theta_H$ = 60° (yellow curve in Fig. \ref{PP}e) deviates from the rest of the dataset, which is presumably due to the change of the magnetic structure at such large angle off the surface normal. The readers may refer to Ref. ~\cite{zhang2020laser} for magnetization hysteresis loops at varying $\theta_H$. The magnetization recovery time constants $\tau_{p1} \approx 400 \pm 150$ ps and $\tau_{p2} \approx 8 \pm 2$ ns, which are similar for different $H_\mathrm{ext}$. The two relaxation processes are quite likely related to energy relaxation by optical and acoustic phonons in semiconductor structures at low temperatures. In addition, Fig. \ref{PP}f plots the corresponding $\theta_M$ as a function of the external magnetic field, which is consistent with the scenario described in \ref{PP}b, i.e., $\theta_M$ increases to $\theta_H$ upon increasing $H_\mathrm{ext}$.

By combining the LLG equation and the Kittel formula, the effective damping coefficient $\alpha_\mathrm{eff}$ is given by the relation 
\begin{equation}
    \alpha_\mathrm{eff} = \frac{2}{\gamma\tau_{T}(H_1+H_2)}
\end{equation}

\noindent Figure \ref{PP}c shows $\alpha_\mathrm{eff}$ as a function of frequency $f$, which is obtained by using $\tau_{T}$ and $\theta_M$ to determine $H_1 = H_\mathrm{ext}cos(\theta_M-\theta_H)+H_Kcos^2\theta_M$ and $H_2=H_\mathrm{res}cos(\theta_M-\theta_H)+H_Kcos(2\theta_M)$. The effective damping coefficient clearly decreases with increasing precession frequency and approaches a constant value. It can be analytically decomposed into an intrinsic damping coefficient ($\alpha_0$) and an extrinsic damping ($\alpha_\mathrm{ext}$). The former is considered to be independent of the external field, while the latter is due to inhomogeneous broadening and depends on $f$. Here, $\alpha_0 = 0.006 \pm 0.002$ for \ce{Cr2Ge2Te6} is lower than that measured for many ferromagnetic materials with perpendicular magnetic anisotropy ~\cite{iihama2014gilbert}. Since \ce{Cr2Ge2Te6} is a semiconductor, this result may be qualitatively understood by considering two facts that are mostly absent in metallic ferromagnets. First, the deviation of the $g$-factor of 2.04 $\pm$ 0.03 from the free-electron value $g = 2$ indicates a tiny orbital contribution to the magnetization. This points to a weak spin–orbit interaction, which induces spin–phonon scattering and thereby strongly suppresses the transverse magnetization relaxation ~\cite{pelzl2003spin}. In addition, spin relaxation due to electron–electron scatterings may be highly suppressed in this layered Cr$_2$Ge$_2$Te$_6$. In contrast, the overall $\alpha_\mathrm{eff}$ obtained in the itinerant ferromagnet Fe$_3$GeTe$_2$ via electron spin resonance is large with an average value of about 0.58 ~\cite{ni2021magnetic}. It is worth noting that even though the laser-induced thermalization can satisfactorily fit the spin relaxation data, this picture based on metallic materials ~\cite{PhysRevLett.76.4250} may not apply to the semiconducting \ce{Cr2Ge2Te6} with a sizable band gap of $\sim$0.2 eV ~\cite{wang2019transition, jiang2020spin, zhuo2021manipulating}. While the electronic structure of \ce{Cr2Ge2Te6} is known to be complicated with defect-induced in-gap states ~\cite{hao2018atomic}, the underlying mechanism of the pump-induced magnetization dynamics demands further studies.

Besides the thermalization effect, the pump-induced magnetization dynamics can be realized through exciton dissociation and charge transfer. In a bilayer CrI$_3$–monolayer WSe$_2$ heterostructure with type-II band alignment, the optical pumping generates exciton in \ce{WSe2}, followed by ultrafast exciton dissociation and charge transfer at the interface. This process leads to an impulsive perturbation to the magnetic interactions in \ce{CrI3} by the hot carriers ~\cite{zhang2020gate}. The magnetization dynamics in CrI$_3$ has been studied using an ultrafast optical pump/magneto-optical Kerr probe technique ~\cite{zhang2020gate}. Figure \ref{PP}e shows the time evolution of the pump-induced Kerr rotation of bilayer CrI$_3$ under an in-plane magnetic field in the range 0-6 T. For all the fields, the MOKE signal shows a sudden change at time zero, followed by an exponential decay on the scale of tens to hundreds of picoseconds. This reflects the incoherent demagnetization process, in which the magnetic order is disturbed instantaneously by the pump pulse and slowly goes back to equilibrium. The frequency of the oscillations can be extracted by fast Fourier transform (FFT, see Fig. \ref{PP}f) of the time traces after subtraction of the exponentially decaying demagnetization dynamics. As shown in Fig. \ref{PP}g, a high-energy ($\omega_h$) and a low-energy ($\omega_l$) mode are identified, which are attributed to the two spin-wave eigenmodes (see inset of Fig. \ref{PP}i) associated with the antiferromagnetic bilayer. Below about 3.3 T, the two initially degenerate modes split. Although both modes soften with increasing field, $\omega_l$ drops nearly to zero frequency. Above 3.3 T, both modes show a linear increase in frequency with a slope equal to the electron gyromagnetic ratio ($\gamma/2 \pi$ = 28 GHz/T). The latter is a characteristic of ferromagnetic resonance in high fields and 3.3 T is a saturation field that is required to fully align the magnetization in-plane. The field-dependence of the spin-wave dynamics can be fully modeled by the coupled LLG equations (see dashed lines in Fig. \ref{PP}g). Here, the low-energy mode $\omega_l$ corresponds to the net moment oscillations along the applied field (the y axis). It drops to zero at the saturation field. The high-energy mode $\omega_h$ corresponds to the net moment oscillations in the x-z plane. Moreover, the damping rate can be evaluated using ($2\pi/\omega\tau$), where $\tau$ is the spin-wave dephasing time. As shown in Fig. \ref{PP}h, the damping is higher below and near saturation. In this regime, the mode frequencies are strongly dependent on internal magnetic interactions, which are sensitive to local inhomogeneity. The exciton dissociation and charge transfer picture is further supported by the control experiment where the spin-wave amplitude strongly depends on the pump wavelength ~\cite{zhang2020gate}. In addition to the exciton physics, this vdW heterostructure exhibits excellent tunability. It has been shown in a dual-gate device that the spin wave frequency can be controlled by electrostatic gating ~\cite{zhang2020gate}.

\subsection{Non-equilibrium spin dynamics near the magnetic critical point}

\begin{figure*}[ht]
\includegraphics[width=1.0\textwidth]{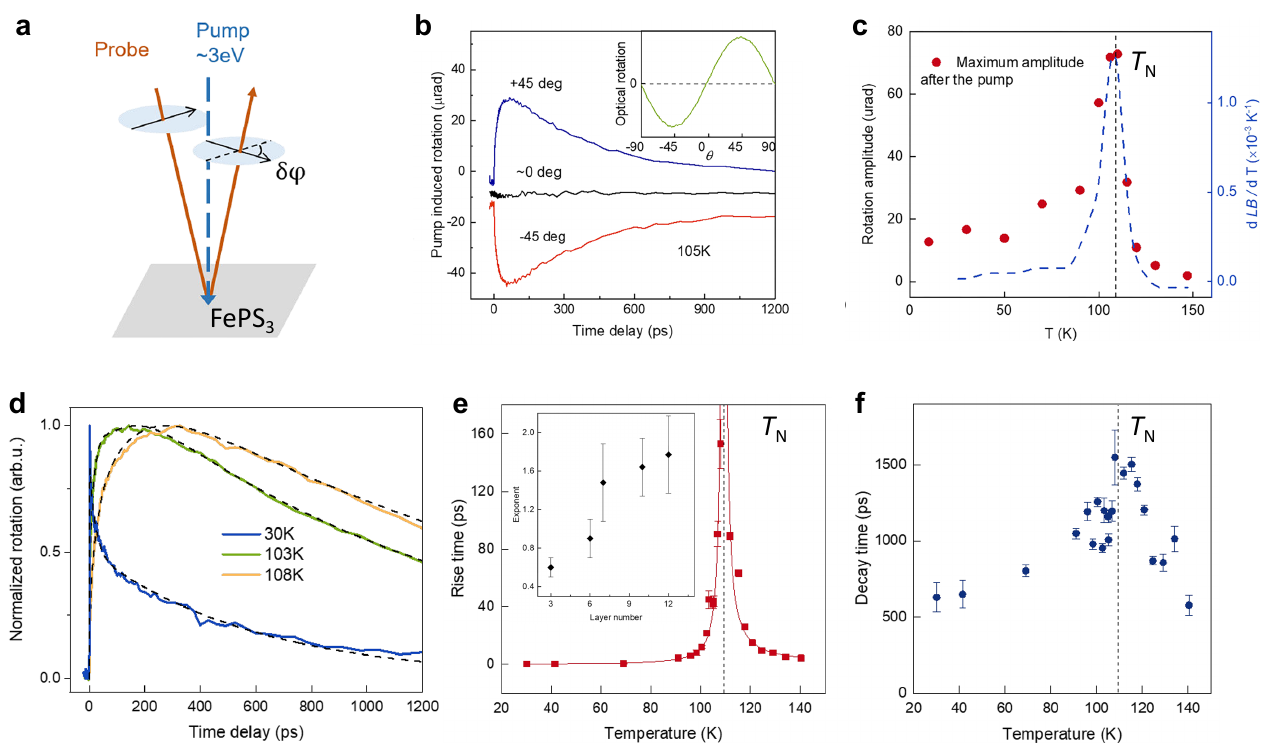}
\caption{\label{MCP}\textbf{a.} Schematics of the pump-probe setup with normal incidence geometry. Pump-induced change in the linear birefringence at 1.6 eV is detected by the polarization rotation angle $\delta\varphi$ of the time-delayed probe beam. \textbf{b.} Time-resolved optical rotation of the probe beam for incident polarization angle of the probe beam $\theta$ = +45° (blue), 0° (black), and -45° (red). The angle $\theta$ is measured from the principal axis of maximum/minimum reflectance. The inset shows that $\delta\varphi$ at 50 ps time delay follows the sin$2\theta$ dependence. The measurement temperature is 105 K, near the Neél temperature ($T_\mathrm{N}$). \textbf{c.} The maximum probe beam rotation $\delta\varphi$ ($\theta$ = 45°) as a function of the sample temperature. For comparison, the blue dashed line is the temperature derivative of the linear birefringence (LB). \textbf{d.} Time-resolved optical rotation of the probe beam $\delta\varphi$ at +45° incident polarization at 30, 105, and 108 K. \textbf{e.} Extracted rise time as a function of temperature. The solid lines are fits using the power-law dependence on the reduced temperature. \textbf{f.} Fitted decay time as a function of temperature. The vertical dashed lines in Panels \textbf{c}, \textbf{e}, and \textbf{f} mark the $T_\mathrm{N}$. All panels are adapted from Ref. ~\cite{zhang2021spin_NL}.}
\end{figure*}

The ultrafast spin dynamics near a magnetic critical point is a challenging problem ~\cite{RevModPhys.49.435, PhysRevLett.123.097601, PhysRevLett.85.1986}. Understanding such nonequilibrium magnetization dynamics will provide important insights into the spin relaxation mechanisms in the materials. Pump-probe magneto-optical spectroscopy has been employed to address this question in atomically thin \ce{FePS3} near its antiferromagnetic critical point ~\cite{zhang2021spin_NL}. As shown in Fig. \ref{MCP}a, \ce{FePS3} sample is optically pumped above its bandgap by $\sim$ 3 eV photons,  creating instantaneous heating by generating hot electrons and phonons. The subsequent time evolution of the linear birefringence is measured by the polarization rotation ($\delta\varphi$) of the time-delayed pulse below the band gap at $\sim$1.6 eV. Figure \ref{MCP}b shows the time evolution of $\delta\varphi$ for three different incident probe polarization angles $\theta$. The pump-induced rotation $\delta\varphi$ shows opposite signs for $\theta$ = $\pm$ 45°. No signal is observed when the probe polarization is aligned with the principal axes ($\theta$ = 0° or 90°). These observations are consistent with pump-induced heating and demagnetization in \ce{FePS3}. In this picture, the pump light heats up the sample and weakens the antiferromagnetic order, which results in a reduction of the linear birefringence and gives rise to a sin$2\theta$ dependence of $\delta\varphi$ (see inset of Fig. \ref{MCP}b). The thermal-induced demagnetization is further supported by the temperature dependence (see Fig. \ref{MCP}c) of the maximum $\delta\varphi$, which reflects the deduction of AF order after the pulse excitation. The temperature dependence of $\delta\varphi$ peaks at $T_N$, at which the magnetic order is most sensitive to small perturbations. The optical pump instantaneously heats up the
sample by $\Delta T$ and changes the optical anisotropy, which can be characterized by the temperature derivative of the linear birefringence ($d\mathrm{LB}/d\mathrm{T}$, blue dashed curve in Fig. \ref{MCP}c).

Remarkably, there is an evident slowdown near $T_N$ in demagnetization. Figure \ref{MCP}d shows the temporal evolution of $\delta\varphi$ (maximum value normalized to 1 for comparison) at three different temperatures. The initial rise in $\delta\varphi$ corresponds to the demagnetization process induced by optical pumping. The subsequent decay is a slow recovery. By fitting the initial rise by exponential growths and the slow recovery by an exponential decay, the obtained temperature dependence of the characteristic rise time $\tau_r$ (the demagnetization time) and the decay time $\tau_d$ are plotted in Fig. \ref{MCP}e, f, respectively. While the increase in decay time near $T_N$ can be attributed to the slower thermal relaxation, the divergent rise time $\tau_r$ directly manifests the criticality at the phase transition point ~\cite{PhysRevLett.56.347, jin2020imaging}. The temperature dependence of rise time can be fitted by a power-law $\tau_r \propto |T/T_N-1|^{-m}$ and obtain an exponent $m \approx 1.7 \pm 0.3$ for $T < T_N$ and $m \approx 1.1 \pm 0.3$ for $T > T_N$ for the 10-layer sample. The different exponent for $T < T_N$ and $T > T_N$ is attributed to the lack of long-range AF order and the emergence of nanoscale domains immediately above $T_N$.

The extracted $m \approx 1.7 \pm 0.3$ for 10-layer \ce{FePS3} samples is in reasonable agreement with the expected correlation time exponent of $\sim$2.1 for a 2D Ising antiferromagnet ~\cite{PhysRevLett.76.4548}. However, the observed decrease in $m$ with decreasing sample thickness (inset of Fig. \ref{MCP}e) cannot be explained by the critical slowing down picture: thinner samples are closer to the ideal 2D limit so that a larger $m$ would be expected. This is because the calculated critical slowing down arises from spontaneous spin fluctuations in thermal equilibrium near the critical point, while optical pumping in this experiment brings the system out of equilibrium. In this process, spin relaxation through coupling to other degrees of freedom (e.g., phonons) becomes important and is beyond the theory of critical slowing down. Further investigation of spin-lattice coupling is recently reported in the XXZ-type vdW antiferromagnet \ce{CoPS3} ~\cite{khusyainov2023ultrafast}.

\begin{figure*}[ht!]
 \centering
 \includegraphics[width=1.0\textwidth]{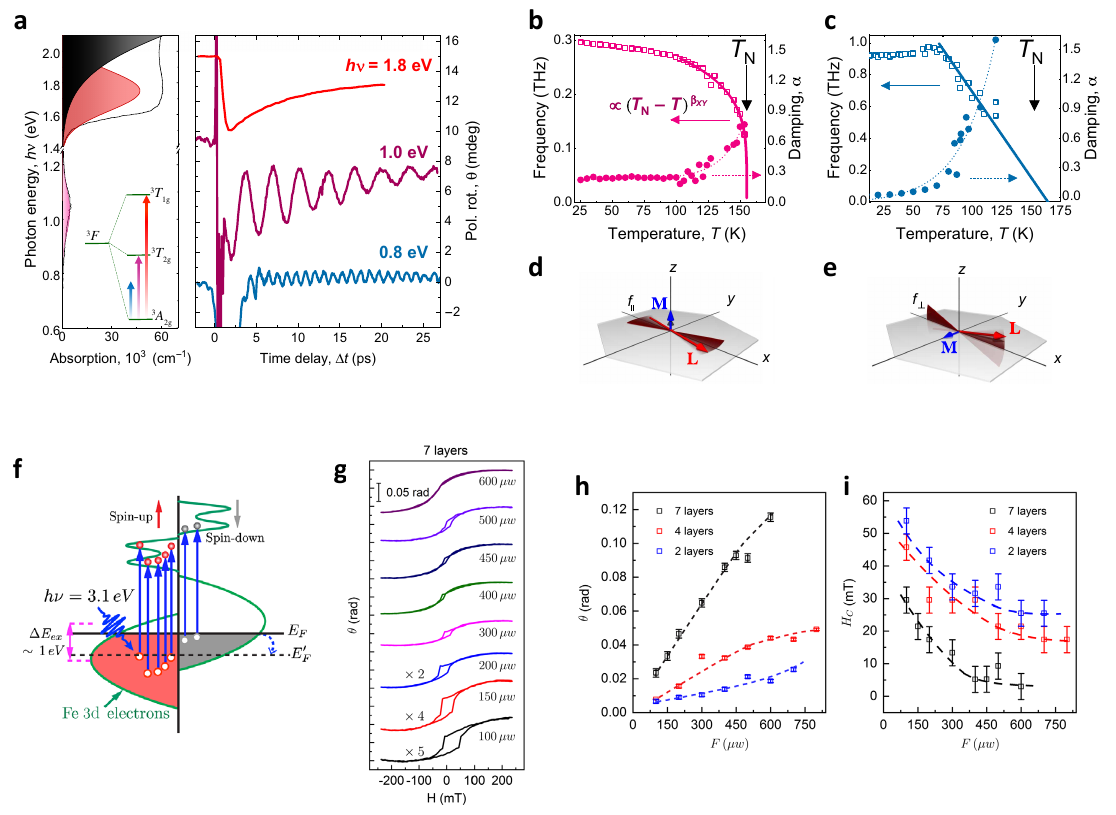}
 \caption{\textbf{a.} Left: Optical absorption spectrum of \ce{NiPS3} displaying the \ce{^{3}A_{2g}} $\rightarrow$ \ce{^{3}T_{2g}} and \ce{^{3}A_{2g}} $\rightarrow$ \ce{^{3}T_{1g}} absorption bands due to the $d-d$ orbital resonances of \ce{Ni^{2+}} ions (in pink and red, respectively) and the onset of the above bandgap absorption due to Ni-S charge transfer transitions (black band). Right: Experimentally detected polarization rotation signal $\theta$ as a function of the delay time $\Delta t$ after excitation with pump pulses at photon energies of 1.8 eV (red), 1.0 eV (pink), and 0.8 eV (blue), respectively. \textbf{b.} and \textbf{c.} Frequency 
 (left axis) and damping factor $\alpha$ (right axis) of the coherent oscillations corresponding to the modes of $f_1$ (pink) and $f_2$ (blue). The solid curves represent the best fit of $(T_N - T)^{\beta}$, which gives $T_N$ = 155 K and $\beta=0.23$, and a linear function to the frequency data.  \textbf{d.} and \textbf{e.} Schematics of the in-plane and out-of-plane magnon modes. \textbf{f.} Schematic of the laser-excited density of states (DOS) in few-layered \ce{Fe3GeTe2} thin films. The photon energy of 3.1 eV causes electron transitions (vertical blue arrows) from occupied states below the Fermi level $E_F$ to the unoccupied states above $E_F$. The simplified DOS diagram is derived from the calculated DOS of the single-layer \ce{Fe3GeTe2} in Ref. [34], where the exchange splitting is estimated to be $\sim$1 eV. \textbf{g.} Excitation fluence dependent ferromagnetism in 7 layers \ce{Fe3GeTe2} films induced and detected by the femtosecond pulsed laser in magneto-optical Kerr effect measurements. Extracted values of the saturated Kerr rotations (\textbf{h}) and coercivities (\textbf{i}) of seven-, four-, and two-layer \ce{Fe3GeTe2} films as a function of the excitation fluence. The dashed curves are guide to the eyes. Panels \textbf{a-e} are adapted from Ref. ~\cite{afanasiev2021controlling}. Panels \textbf{f-i} are adapted from Ref. ~\cite{PhysRevLett.125.267205}.}
 \label{Light_control}
\end{figure*}

\subsection{Light control of magnetism}

The intimate connection between long-range magnetic order, low dimensionality, and the search for energy-efficient driving forces for the magnetic control of 2D vdW materials raised a fundamental question: How can we tune the magnetic features of 2D vdW magnets via laser-probes? It has been previously demonstrated that short laser excitations can control the magnetic features of thin films included in all-optical switching (AOS) devices~\cite{Rasing07,rasing_rmp2010}. Such approach provides a feasible pathway for energy-efficient, faster, and low-power magneto-optical implementations as achieved on different non-vdW compounds~\cite{Fullerton14,Lambert14,Koopmans19,Rasing07}. Moreover, AOS offers a broad range of applications on data memory technologies~\cite{rasing_rmp2010}, and its integration in spintronics would pave the way to the emerging field of ultrafast spin-optronic devices. In this context, optical pumping of the electronic transitions toward the higher-level orbital states, that is, orbital resonances provides the most direct access to the admixing and subsequent control of important magnetic properties, such as the magnetic anisotropy ~\cite{afanasiev2021controlling}.

Optical control of magnetism has been demonstrated in a few materials such as CrI$_3$\cite{Hicken22}, CrCl$_3$\cite{Strungaru22}, CrGeTe$_3$\cite{Hicken23cc,Johansson23} and \ce{NiPS3}\cite{afanasiev2021controlling}, which is a vdW layered magnet with XY-type antiferromagnetism. As shown in Fig. \ref{Light_control}a, the orbital resonances in \ce{NiPS3} correspond to a pair of $d-d$ transitions \ce{^{3}A_{2g}} $\rightarrow$ \ce{^{3}T_{1g}} (1.73 eV) and \ce{^{3}A_{2g}} $\rightarrow$ \ce{^{3}T_{2g}} (1.07 eV) emerging within the \ce{^{3}F} ground state multiplet of the \ce{Ni^{2+}} ion split by the octahedral crystal field. By pumping the sample at 10 K, well below $T_N$ using linearly polarized ultrashort pulses, the time-resolved polarization rotation $\theta$ reveals a notable sensitivity to the photon energy of the excitation. When excited at the \ce{^{3}A_{2g}} $\rightarrow$ \ce{^{3}T_{2g}} resonance ($h\nu$ = 1.0 eV), $\theta$ displays a damped oscillation as a function of the pump-probe time delay $\Delta t$, with a frequency of $f_1$ = 0.30 THz. No coherent oscillations were observed when excited at the \ce{^{3}A_{2g}} $\rightarrow$ \ce{^{3}T_{1g}} resonance ($h\nu$ = 1.8 eV). Detuning the photon energy below the absorption lines of the resonances ($h\nu$ = 0.8 eV) reveals another higher-frequency mode at $f_2$ = 0.92 THz. The frequencies of $f_{1,2}$ have no match in the phonon spectrum of \ce{NiPS3} ~\cite{kim2019suppression}, indicating their magnetic origin. On the other hand, the excitation of $f_1$ mode shows pronounced resonance with the \ce{^{3}A_{2g}} $\rightarrow$ \ce{^{3}T_{2g}} transition, while $f_2$ mode exhibits the off-resonant character as it can be excited in a broad window of optical pumping.

The temperature dependence of the $f_{1,2}$ modes further evidences their sensitivity to the magnetic ordering. As temperature increases towards $T_\mathrm{N}$, the frequency of $f_1$ mode shows critical softening that can be fitted to a power law $(T_\mathrm{N}-T)^{\beta}$ (see Fig. \ref{Light_control}b), with $T_\mathrm{N}$ = 155 K and a critical exponent $\beta = 0.23$. The values of $T_\mathrm{N}$ and $\beta$ match with the literature data remarkably well ~\cite{bramwell1993magnetization, taroni2008universal}. In contrast, the softening of $f_2$ mode is linear with an extrapolation of complete softening at 170 K. Based on a phenomenological model, two magnon modes are expected in a compensated antiferromagnet. The frequency of $f_1$ mode agrees well with the in-plane magnon ($f_{\parallel}$ = 280 GHz, see Fig. \ref{Light_control}d) with a 2D critical scaling. Presumably, the higher-frequency oscillation at $f_2$ can be assigned to the out-of-plane magnon ($f_{\perp}$, see Fig. \ref{Light_control}e). However, these assumptions do not agree with the recently reported values for the zone-center magnons at the significantly higher frequencies of 1.69 ~\cite{PhysRevB.98.134414} and 2.4 THz ~\cite{kang2020coherent}. Time-resolved measurements in higher magnetic fields are required to unambiguously establish the origin of the coherent mode $f_2$.

Magnetization and magnetic anisotropy can also be tuned via tailoring the electronic structure. The itinerant magnetic order in \ce{Fe3GeTe2} was demonstrated to satisfy the Stoner criteria $ID(E_F) > 1$, where $I$ is the Stoner parameter, describing the exchange energy between electrons with up and down spins and $D(E_F)$ represents the density of states near the Fermi level ~\cite{PhysRevB.93.134407, tan2018hard}. Figure \ref{Light_control}f) shows a simplified density of states diagram of the few-layered \ce{Fe3GeTe2} films, which is derived from the calculated DOS of single-layer \ce{Fe3GeTe2} ~\cite{PhysRevB.93.134407}. The schematic DOS is composed of 3$d$ electrons (green lines) that mainly provide the magnetic moments. After excitations by femtosecond laser pulses with 3.1 eV photon energy, the majority (red) and minority electrons (gray) below $E_F$ are excited to the unoccupied states, leaving excited holes near $E_F$. The redistributed electronic states would shift $E_F$ down to $E'_F$ (black dashed line), thus enhancing the DOS near the Fermi level. This leads to the Stoner instability and thus strengthens the ferromagnetic order that manifested as the enhancement of $T_C$ ~\cite{PhysRevLett.125.267205}.

In Fig. \ref{Light_control}g, under excitations of femtosecond laser pulses, the Kerr rotations ($\theta$) of seven-layer \ce{Fe3GeTe2} samples show clear magnetic hysteresis loops at room temperature under out-of-plane magnetic field. By increasing the excitation intensity from 100 to 800 $\mu$W, the hysteresis loops are significantly modified with distinct variations of both Kerr rotations and coercivities ($H_C$). As the intrinsic $T_C$ of the seven-layer \ce{Fe3GeTe2} is $\sim$200 K, the obtained magnetic hysteresis loops at room temperature demonstrate the emergence of room-temperature ferromagnetism. The extracted saturated Kerr rotations and coercivities of the hysteresis loops of \ce{Fe3GeTe2} samples with different thicknesses are summarized in Figs. \ref{Light_control}h and \ref{Light_control}i, respectively, as a function of the laser fluence. For all three different layer \ce{Fe3GeTe2} samples, the saturation magnetization characterized by the saturated Kerr rotations increases with laser fluence, which is consistent with the enlarged difference between the majority and minority electron density upon optical excitation. The easy axis coercivity in \ce{Fe3GeTe2} is expected to be proportional to the perpendicular magnetic anisotropy. The decreasing magnetic anisotropy under optical excitations is attributed to the shift of the spin-orbital-coupling-induced bands away from the Fermi level, which echoes the suppression of magnetic anisotropy in \ce{Fe_{3-x}GeTe2} caused by the hole doping ~\cite{park2019controlling}.

\subsection{Collective excitations}

\begin{figure*}
\includegraphics[width=0.76\textwidth]{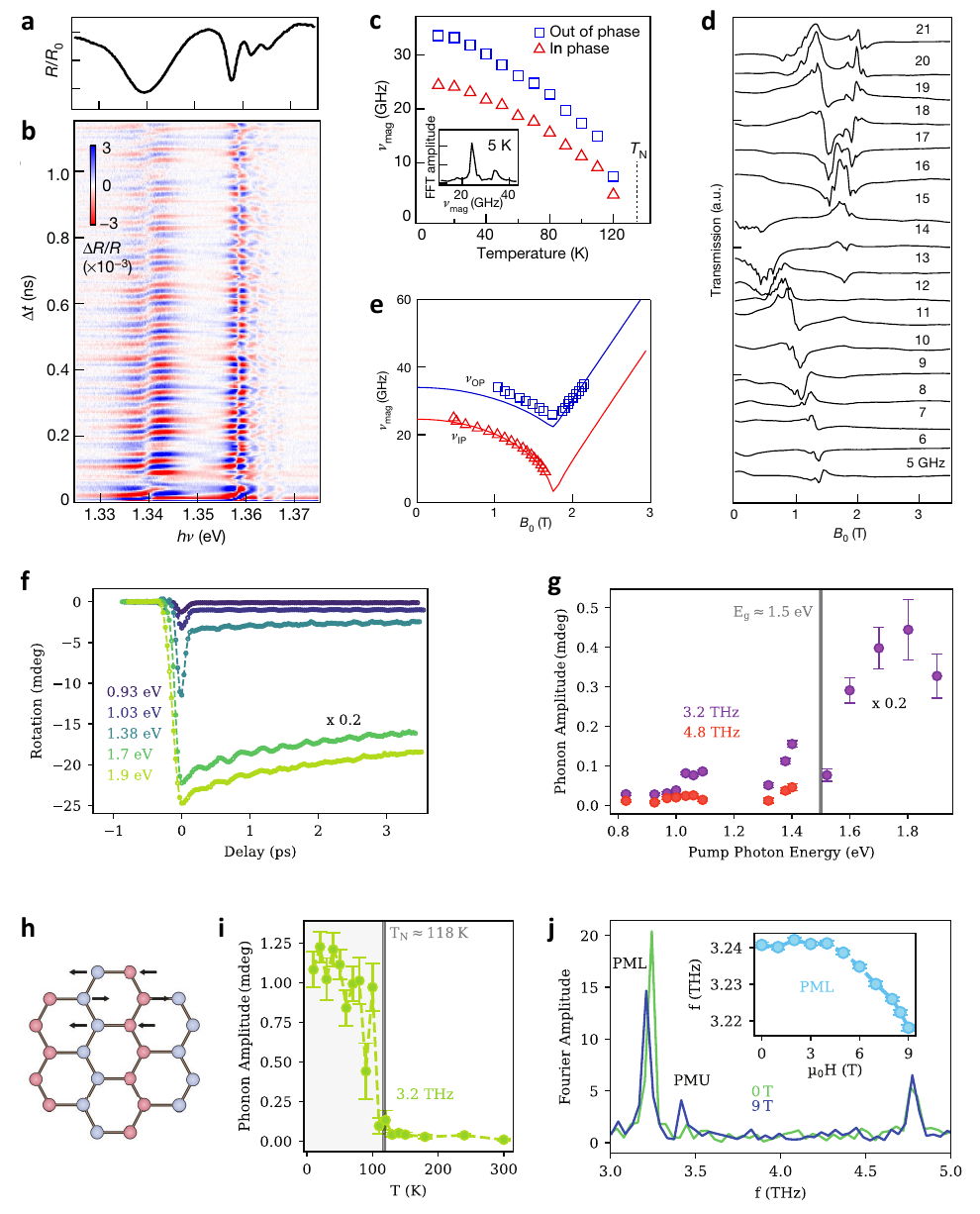}
\caption{\label{ex} \textbf{a.} Reflectance ($R$) from CrSBr normalized to that of the SiO$_2$ substrate ($R_0$).\textbf{b.} Transient reflectance spectra $\Delta R/R$ as a function of pump–probe delay ($\Delta t$) and probe photon energy ($h\nu$). An incoherent background has been subtracted from $\Delta R/R$. The pseudo-color scale is $\Delta R/R$, where $R$ is reflectance without pump and $\Delta R$ is the pump-induced change in reflectance. \textbf{c.} The two spin-wave frequencies are shown as a function of sample temperature below $T_N$. The inset is the probe $h\nu$-integrated FFT trace showing the two peaks at 24 GHz and 34 GHz at 5 K. \textbf{d.} Magnetic resonance spectra at the indicated frequencies (5-21 GHz) from a bulk CrSBr crystal at $T$ = 5 K, with the magnetic field applied along the $c$ axis. The spectra are offset for clarity. \textbf{e.} Peak resonance frequencies as a function of magnetic field (symbols). The solid curves are dispersion fits from linear spin-wave theory. \textbf{f.} Transient rotation of the polarization measured from an exfoliated \ce{FePS3} flake at 10 K for different pump photon energies. The pump and probe beams were linearly polarized 45° away from each other. The pump fluence is kept constant at 2 $\mathrm{mJ}\cdot \mathrm{cm}^{-2}$ for pump photon energy below the band gap and to 1 $\mathrm{mJ}\cdot \mathrm{cm}^{-2}$ for pump photon energy above the band gap. \textbf{g.} Amplitude of the 3.2 THz (purple) and the 4.8 THz (orange) phonon modes as a function of the pump photon energy. \textbf{h.} Schematic representation of the Fe-ions lattice in \ce{FePS3}, forming a magnetic zig-zag pattern. The arrows indicate exemplarily the motion of the ions in the Raman-active phonon mode of 3.2 THz. \textbf{i.} Amplitude of the 3.2 THz mode as a function of the sample temperature. \textbf{j.} Fast Fourier transformation of the extended time-traces with the absence of an external field (green) and at 9 T (blue). PML labels the peak of the lower branch of the phonon-magnon mode, which corresponds to the 3.2 THz phonon when no external field is applied. The peak labeled PMU is the upper branch phonon–magnon mode, appearing at 9 T. The other peak corresponds to the 4.8 THz phonon mode, which is unaffected by the external magnetic field. The inset shows the frequency shift of the PML mode caused by the phonon–magnon hybridization. Panels \textbf{a-d} are adapted from Ref.~\cite{bae2022exciton}. Panels \textbf{f-j} are adapted from Ref.~\cite{mertens2023ultrafast}.}
\end{figure*}

Collective excitations are ubiquitous in condensed matter. Their existence in magnetic materials gives rise to intriguing light-matter interactions. Driving the collective excitations with light thus offers a promising platform for realizing unconventional many-body phenomena and phases.

Atomically thin flakes of CrSBr maintain the bulk magnetic structure down to the ferromagnetic monolayer with a Curie temperature $T_C$ = 146 K and to the antiferromagnetic bilayer with a N$\mathrm{\acute{e}}$el temperature $T_N$ = 140 K ~\cite{lee2021magnetic}. CrSBr is also a direct-gap semiconductor down to the monolayer, with an electronic gap of 1.5 eV and an excitonic gap of 1.34 eV ~\cite{telford2020layered}. The coexistence of both magnetic and semiconducting properties implies that a spin wave may coherently modulate the electronic structure. As a result, such exciton-magnon coupling allows the launch and detection of spin waves from strong absorption, emission, or reflection of light in the energy range corresponding to excitonic transitions. Here, we clarify that distinct from the exciton-magnon coupling in CrSBr, the term ``exciton-magnon" in the literature ~\cite{PhysRevLett.15.654, PhysRevLett.15.1023, moriya1968theory, tanabe1982excitons} specifically describes the magnon-induced electric dipole transition at the ion site in a magnetic material. Recently, such exciton-magnon transition is shown to be responsible for the photo-induced magnetoelastic interaction ~\cite{PhysRevLett.127.077202} and magnetoelectricity ~\cite{bossini2018femtosecond}. Remarkably, in vdW correlated insulator \ce{NiPS3}, spin–orbit-entangled exciton transition leads to a transient metallic state that preserves long-range antiferromagnetism ~\cite{belvin2021exciton}.

Based on first-order perturbation theory, the shift in the exciton energy ($\Delta E_\mathrm{ex}$) owing to changes in the interlayer electron-exchange interaction can be evaluated by $\Delta E_\mathrm{ex} \propto \mathrm{cos^2}(\theta /2)$ where $\theta$ is the angle between the magnetic moments in neighboring layers ~\cite{wilson2021interlayer}. In the AFM state ($\theta = \pi$), the interlayer hybridization is spin forbidden and $\Delta E_\mathrm{ex} = 0$; in the FM state ($\theta = 0$), the interlayer electron-exchange interaction is the greatest and $\Delta E_\mathrm{ex}$ is -20 meV ~\cite{wilson2021interlayer}. The dependence of $\Delta E_\mathrm{ex}$ on $\theta$ is the basis for exciton sensing of coherent spin waves ~\cite{bae2022exciton}. 

To investigate the dynamical change in $\Delta E_\mathrm{ex}$, CrSBr is excited by a femtosecond laser pulse with above-gap photon energy ($h\nu1 = 1.7$ eV) and the change in reflectance at a variable time delay ($\Delta t$) is measured by a broadband probe pulse ($h\nu2 = 1.3-1.4$ eV) ~\cite{bae2022exciton}. As shown by the static reflectance spectrum in Fig. \ref{ex}a, the peaks in $\Delta R/R$ correspond to excitonic transitions. Upon photoexcitation, the strong oscillatory components in Fig. \ref{ex}b come from the coherent magnons and the coupling between excitons and coherent magnons is revealed by clear $\pi$-phase flips of the oscillatory signal at the excitonic transitions ~\cite{luer2009coherent, kumar2001investigations}. The frequencies of the coherent magnons can be determined by an FFT of the oscillatory signal, yielding $\nu_\mathrm{mag1} = 24.6 \pm 0.7$ GHz and $\nu_\mathrm{mag2} = 34 \pm 1$ GHz (see inset of Fig. \ref{ex}c for a typical FFT spectrum at $T$ = 5 K). With increasing temperature, the magnetic order decreases and this results in lowering of the spin-wave frequencies (Fig. \ref{ex}c). Around $T_N$, both $\nu_\mathrm{mag1}$ and $\nu_\mathrm{mag2}$ approach zero. The temperature dependence of $\nu_\mathrm{mag1}$ and $\nu_\mathrm{mag2}$ closely follows that of the magnetic-order parameter ~\cite{lee2021magnetic}.

The assignments of coherent magnon modes from excitonic sensing are supported by magnetic resonance spectroscopy. Figure \ref{ex}d shows a series of magnetic resonance spectra at selected microwave frequencies ($\nu$ = 5–21 GHz) for bulk CrSBr at $T$ = 5 K. The spectra reveal a single resonance in the low-frequency ($\leq$18 GHz) region and two resonances in the high-frequency ($>$22 GHz) region. As shown in Fig. \ref{ex}e, the peak frequencies of the resonances are plotted as a function of the magnetic field applied along the $c$ axis ($B_0$). For $B_0$ smaller than a saturation field ($B_\mathrm{sat} \approx$ 1.7 T), two branches of magnons are observed whose frequencies decrease with increasing $B_0$, consistent with the reduction in AFM order as spins are progressively canted away from the easy $b$ axis. The frequencies of these two branches are assigned to the in-phase ($\nu_\mathrm{IP}$) and out-of-phase ($\nu_\mathrm{OP}$) spin precessions, similar to those observed in the 2D AFM materials of chromium chloride and chromium iodide ~\cite{macneill2019gigahertz, zhang2020gate}. Above $B_\mathrm{sat}$, the spins are fully polarized parallel to $B_0$, and $\nu_\mathrm{mag}$ increases linearly with $B_0$, which is expected for a ferromagnetic resonance. The magnetic resonance spectroscopy data can be fitted to a linear spin-wave theory (solid curves in Fig. \ref{ex}e) ~\cite{scheie2022spin}.

The laser-driven THz phonon hybridized with a magnon mode is reported in the antiferromagnetic vdW semiconductor \ce{FePS3} ~\cite{mertens2023ultrafast}. Figure \ref{ex}f shows the transient polarization rotation measured with different pump photon energies with sample temperature at 10 K, well below $T_N$ =118 K. The Fourier transform reveals a superposition of two coherent oscillations with frequencies of 3.2 THz and 4.8 THz, which match the eigenfrequencies of Raman-active optical phonons reported in the literature ~\cite{lee2016ising, wang2016raman, mccreary2020quasi, PhysRevB.103.064431}. The coherent oscillations can be ascribed to changes in either the linear crystallographic birefringence or dichroism, induced by the phonon-modulated magnetic zigzag pattern (see Fig. \ref{ex}h). The spectral dependence of the phonon amplitude is summarized in Fig. \ref{ex}g. The extracted amplitude of the phonon modes increases in the presence of electronic transitions, i.e., in the region $\sim$1.1 eV ($d–d$ transitions) and then above 1.5 eV (the energy of the band gap), where a steep slope related to the onset of the band-gap is observed. The phonon modes can be induced and detected only in the presence of the long-range antiferromagnetic order. Figure \ref{ex}i shows the temperature dependence of the amplitude of the 3.2 THz phonon mode, which vanishes above the Néel temperature.

By applying an external magnetic field up to 9 T, the 3.2 THz phonon hybridizes to a zone-center magnon. In the 9 T data shown in Figure \ref{ex}j, the original 3.2 THz mode shift towards lower frequency, and a new mode appears at 3.42 THz. These observations can be explained by considering that, in the presence of an external magnetic field, the phonon mode at 3.2 THz hybridizes with a magnon mode with an eigenfrequency of 3.6 THz at 0T ~\cite{mccreary2020quasi, PhysRevB.103.064431} giving rise to two phonon–magnon branches ~\cite{PhysRevB.104.134437, PhysRevLett.127.097401, zhang2021coherent}: the phonon–magnon lower branch (PML) and the phonon–magnon upper branch (PMU).

\section{Current-induced spin-torque resonance}

Spin current injected into a ferromagnetic material produces a spin-transfer torque (STT) ~\cite{ralph2008spin}, which is essential for low-power manipulation of magnetization in spintronic devices, such as magnetic random-access memory, race-track memory, and logic circuit ~\cite{chen2010advances, kawahara2012spin, thomas2014perpendicular, locatelli2014spin, fong2015spin, parkin2015memory}. Recently, many studies have been reported on the STT of spin currents generated by the spin Hall effect ~\cite{hirsch1999spin, kato2004observation, valenzuela2006direct, kimura2007room, sinova2015spin} and the Rashba-Edelstein effect ~\cite{edelstein1990spin, ganichev2002spin, sanchez2013spin}, where momentum locking due to spin-orbit coupling (SOC) plays a significant role in the spin-current generation ~\cite{manchon2015new}. In particular, the emergence of vdW magnets has stimulated growing interest in the exotic spin transport phenomena in spintronic devices. One of the major tasks of these studies is to evaluate the spin Hall angle, which represents the conversion ratio of the generated spin current to the injected charge current ~\cite{hoffmann2013spin, horaguchi2020highly}.

\begin{figure*}[ht]
\includegraphics[width=1.0\textwidth]{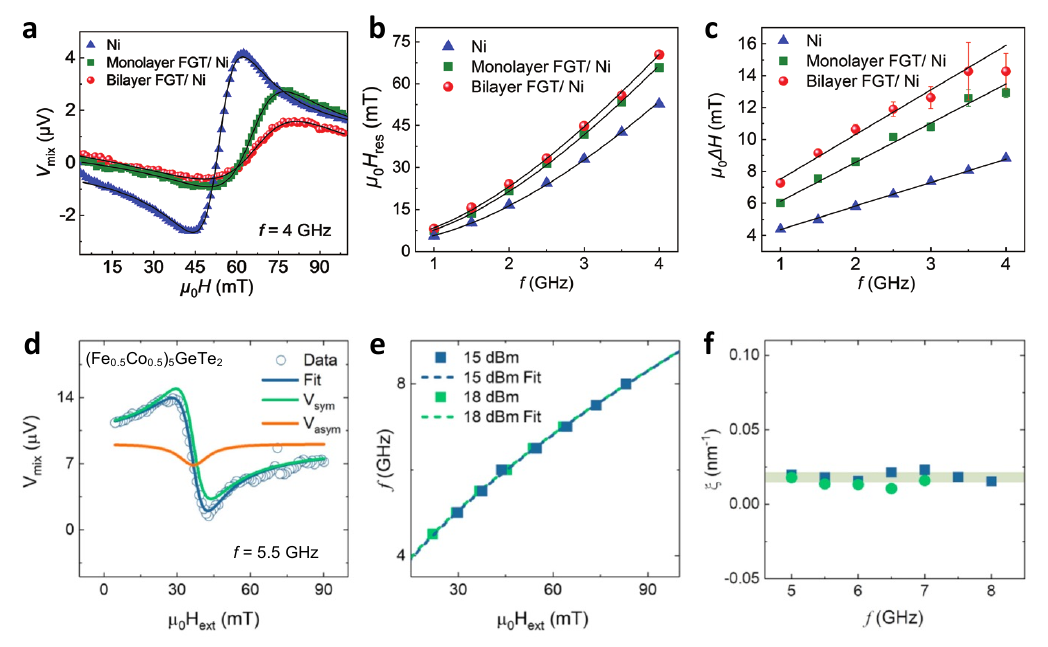}
\caption{\label{STFMR}\textbf{a.} Typical ST-FMR spectra of Ni (blue), monolayer-FGT/Ni (green) and bilayer-FGT/Ni (red) recorded at 4 GHz at room temperature. Frequency dependence of \textbf{b.} the resonance field and \textbf{c.} the linewidth of these devices. \textbf{d.} The ST-FMR signal of an FCGT/Co$_{0.9}$ Fe$_{0.1}$ sample at 5.5 GHz, 18 dBm. The solid lines are fits that show the symmetric (orange) and antisymmetric (yellow) Lorentzian contribution. \textbf{e.} ST-FMR frequency as a function of resonance field; the blue (15 dBm) and green (18 dBm) line is a fit to the Kittel formula. \textbf{f.} Charge to spin conversion coefficient ($\xi$) as a function of frequency at 15 and 18 dBm. Panels \textbf{a-c} are adapted from Ref. ~\cite{chen2021proximity}. Panels \textbf{d-f} are adapted from Ref. ~\cite{PhysRevMaterials.6.044403}.}
\end{figure*}

ST-FMR has been employed to study the proximity effect in a device that consists of a Fe$_3$GeTe$_2$ (FGT) flake and Ni thin films ~\cite{chen2021proximity}. Figure \ref{STFMR}c shows the typical ST-FMR spectra recorded from three series of devices: Ni, monolayer-FGT/Ni and bilayer-FGT/Ni. The symmetric components of these spectra are close to zero, indicating the negligible spin-orbit torques induced by Au contact pads. The spectra demonstrate the modulation of Ni magnetism by FGT, evident by the increasing resonance field ($H_\mathrm{res}$) and linewidth ($\Delta H$) when Ni is integrated with monolayer and bilayer FGT. $H_\mathrm{res}$ as a function of frequency are shown in Fig. \ref{STFMR}d. Based on the fitting to a Kittel equation, one can identify a decreasing $M_\mathrm{eff}$ going from the Ni device to the monolayer-FGT/Ni and bilayer-FGT/Ni devices. The decreased $M_\mathrm{eff}$ value in FGT/Ni may be attributed to the increase of the perpendicular magnetic anisotropy (PMA) in Ni. Apart from the interfacial $d$--$d$ hybridization, the increased PMA of Ni may be associated with the interlayer magnetic exchange coupling between the Fe and Ni atoms.

Moreover, the FGT interface also leads to enhanced magnetic damping in Ni, as indicated by the frequency dependence of and $\Delta H$ shown in Fig. \ref{STFMR}e. The frequency-dependent linewidths can be well fitted using
\begin{equation}
    \Delta H = \Delta H_0 + \frac{4\pi\alpha}{\gamma}f
\end{equation}

\noindent where $\Delta H_0$ is the linewidth at zero frequency determined by inhomogeneous broadening, and $\alpha$ is the Gilbert damping of Ni. The increased inhomogeneous broadening may result from interfacial magnetic coupling, which gives rise to the inhomogeneous magnetization texture ~\cite{chen2021proximity,xie2020characterizing}. The enhanced Gilbert damping suggests additional loss of the spin angular momentum, which may be attributed to the spin pumping effect and the exchange coupling between the Fe and Ni atoms.

Another significant work is ST-FMR measurements of (Fe$_{0.5}$Co$_{0.5}$)$_5$GeTe$_2$ (FCGT, 21 nm, $\sim$11 unit cells) on Co$_{0.9}$ Fe$_{0.1}$ (12 nm) ~\cite{PhysRevMaterials.6.044403}. AA$^{\prime}$-stacked FCGT is a wurtzite-structure polar magnetic metal ($T_c \sim$ 350 K) with broken spatial inversion and time reversal symmetries. It exhibits a N$\acute{\mathrm{e}}$el-type skyrmion lattice as well as a Rashba-Edelstein effect at room temperature. A typical ST-FMR signal for an AA$^{\prime}$-FCGT/Co$_{0.9}$ Fe$_{0.1}$ is shown in Fig. \ref{STFMR}f. The spectrum can be fitted well to a sum of symmetric and asymmetric components, and ST-FMR frequency as a function of resonant field (see Fig. \ref{STFMR}g) is in good agreement with the Kittel formula. The charge-to-spin conversion coefficient ($\xi$),represented by $\theta_\textrm{SH}$ in Eq. (15), as a function of frequency is essentially constant, as shown in Fig. \ref{STFMR}h. The average $\xi$ of the AA$^{\prime}$-FCGT/Co$_{0.9}$ Fe$_{0.1}$ system is $\sim$0.017 $\pm$ 0.003 nm$^{-1}$. The ST-FMR signal of the AA$^{\prime}$-FCGT/Co$_{0.9}$ Fe$_{0.1}$ sample may originate from the Rashba-Edelstein effect of the Te/Co$_{0.9}$ Fe$_{0.1}$ interface or the spin Hall effect from the AA$^{\prime}$-FCGT system. Even though the second-harmonic Hall signal was observed in the single-layer AA$^{\prime}$-FCGT, more approaches are still needed to distinguish the contributions of the ST-FMR signal in detail in future work.

\begin{table*}[ht]
\textbf{\caption{Comparison between different FMR techniques and methodologies.}}
\vspace{5pt}
\centering
\renewcommand{\arraystretch}{1.7}
\begin{tabular}{|p{2cm}||p{4cm}|p{3.2cm}|p{3.75cm}|p{3.75cm}|}
\hline
\textbf{Technique}  & \textbf{Spin excitation methods}     &   \textbf{Detection methods}   &  \textbf{Sample requirements}     &  \textbf{Unique capabilities}     \\
\hline
Broadband FMR  & Microwave injected through co-planar waveguide & Microwave absorption  &  Bulk crystals, thin films & It has high flexibility and is compatible with cavity resonator to enhance the sensitivity    \\

Optical FMR  & Optical pump using ultrafast laser pulses   & Magneto-optical Faraday/Kerr effect, transient reflectivity, x-ray magnetic circular dichroism  & Bulk crystals, thin films, exfoliated thin flakes &  It has spatial resolution for mapping domain structures and time-resolution for measuring spin relaxation time. \\

Spin-torque FMR  & Microwave current injected through on-chip coplanar waveguide pattern  & DC voltage generated from the rectification between AC current and resistance oscillation  &  Heavy metal/ferromagnet bilayer structures, exfoliated thin flakes,  & It provides direct access to the spin-torque induced spin-charge conversion  \\
 \hline

\end{tabular}
\end{table*}

\section{Recent advances in FMR techniques }

Despite the unique capabilities of the broadband FMR, optical FMR, and spin-torque FMR techniques as summarized in \textbf{Table II}, each technique has its limitations. Therefore, we provide the caveats as follows when studying spin dynamics using these techniques. The sensitivity of broadband FMR is typically limited by the size of the antenna. This technique thus is more effective when probing the large-size thin films of vdW materials in comparison with the exfoliated flakes. The optical FMR requires synchronization with microwave signals which demands an elaborate setup of optical components. In addition, it faces challenges when measuring samples with a weak magneto-optical effect. The spin-torque FMR experiments entail a complex and intricate fabrication process, which hinders its scalability and practicality for high-throughput measurements.

Looking forward, we envision many emergent spin dynamics probes will be applied to study 2D vdW magnets. Here, we highlight several recent technical advances in probing phase-resolved magnetic dynamics, magnon-phonon hybridization, local probe of spin transport, and element-specific magnetization dynamics. In addition to unraveling fundamental properties, these techniques involving fiber-based magneto-optical detection, superconducting resonator, and nitrogen-vacancy magnetometry also represent promising applications in hybrid quantum magnonic devices and quantum sensing.

\begin{figure*}[ht!]
 \centering
 \includegraphics[width=1.0\textwidth]{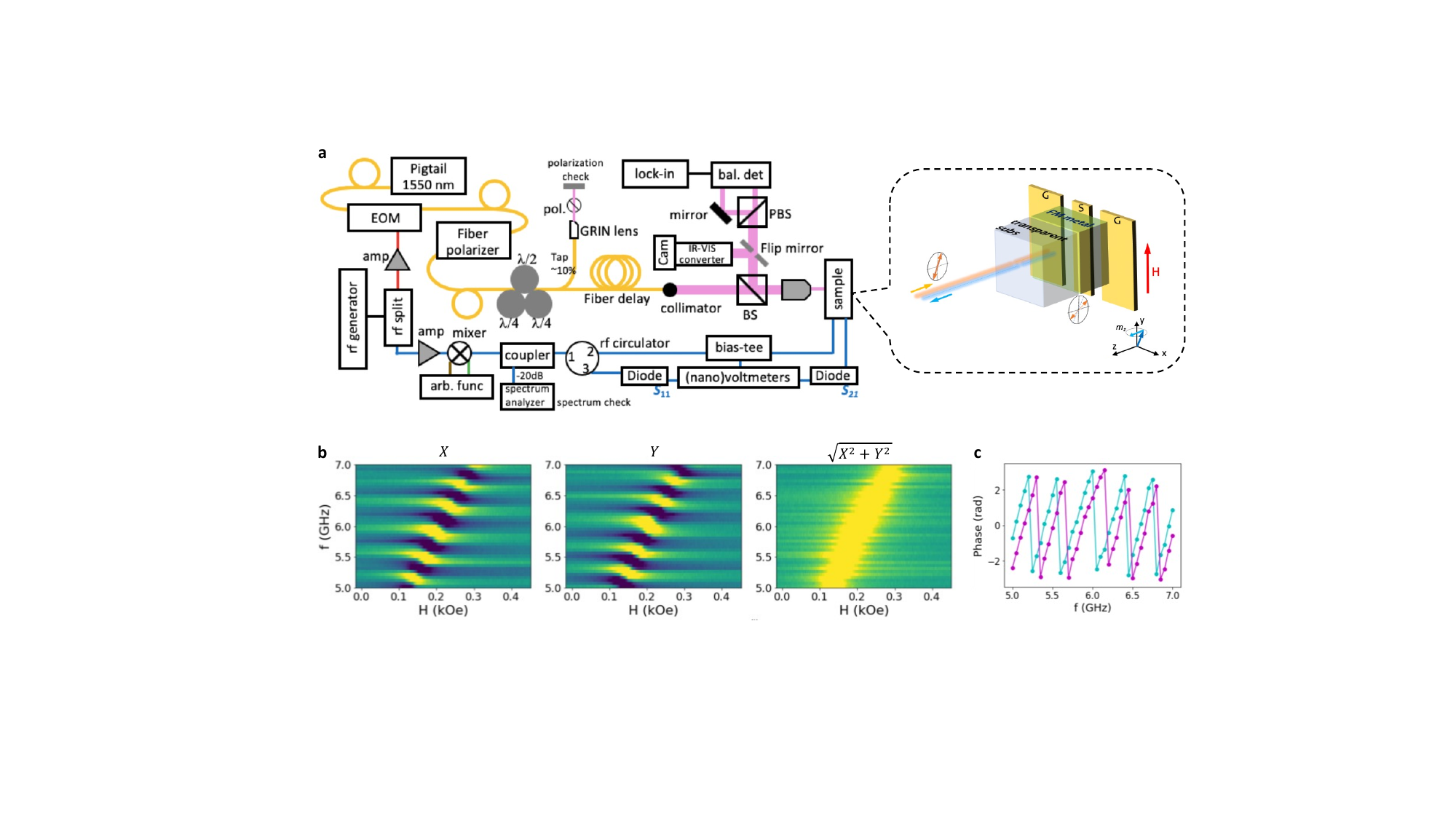}
 \caption{\textbf{a.} Schematic of the optical detection of phase-resolved magnetization dynamics setup. After the rf splitter, the optical path (upper part) contains amplifier, 1550-nm infrared laser module, electro-optic modulator (EOM), fiber polarizer, fiber polarization controller, beam splitter (BS) and focusing lens; the electrical path (lower part) contains amplifier, mixer, coupler, spectrum analyzer, diodes, circulator, bias-tee, and nanovoltmeter. (PBS = polarizing beam splitter, Cam = camera, bal.det = balanced detector, arb. func = arbitrary waveform generator.) The dashed box shows the detecting mechanism for the dynamic Kerr effect. The applied DC magnetic field is parallel to the ground-signal-ground (G-S-G) lines of the coplanar waveguide. \textbf{b.} The intensity map of the FMR scan (5.0--7.0 GHz) for the Fe/Pt bilayer of the in-phase ($X$) and quadrature ($Y$) components, and total amplitude ($\sqrt{X^2+Y^2}$) as a function of the magnetic field and frequency. \textbf{c.} The phase evolution as a function of frequency. All panels are adapted from Ref. ~\cite{xiong2020detecting}.}
 \label{MOD}
\end{figure*}

\subsection{Magneto-optical detection of phase-resolved ferromagnetic resonance}

\begin{figure*}[hb]
 \centering
 \includegraphics[width=1.0\textwidth]{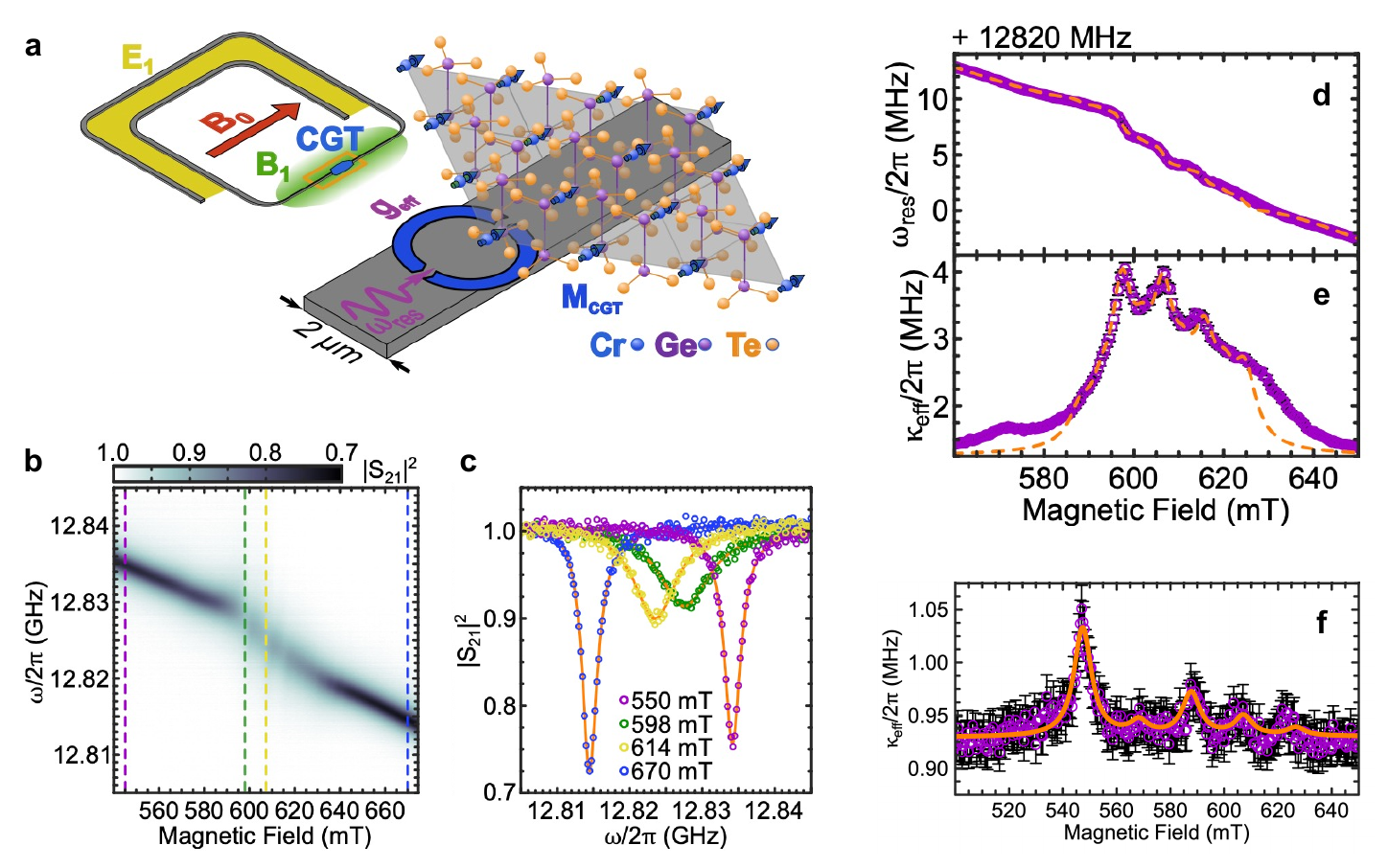}
 \caption{\textbf{a.} Schematic of a superconducting resonator indicating the areas of high $E_1$-field (yellow) and $B_1$-field (green) intensities, as well as the orientation of the externally applied field $B_0$. A zoom-in of the section loaded with a Cr$_2$Ge$_2$Te$_6$ flake is shown. \textbf{b.} $|S_{21}|^2$ as a function static magnetic field $B_0$ and frequency, with the microwave transmission encoded in the color. \textbf{c.} $|S_{21}|^2$ as a function of frequency at fixed magnetic fields, indicated in \textbf{b} by dashed vertical lines. \textbf{d} and \textbf{e.} Resonance frequency $\omega_{res}$ and effective loss rate $\kappa_\mathrm{eff}$ as a function of magnetic field. The dashed orange lines are the results from the semi-optimized fit. \textbf{f.} Effective loss rate $\kappa_\mathrm{eff}$ as a function of the magnetic field of a resonator loaded with the thinnest CGT sample ($\sim 11 nm$). All panels are adapted from Ref. ~\cite{Zollitsch_arXiv2022}.}
 \label{MPC}
\end{figure*}

The development in quantum magnonics ~\cite{tabuchi2016quantum, lachance2019hybrid, yuan2022quantum} highlight the needs for detecting spatial- and phase-resolved magnetization dynamics adaptable to micro- and nano-scale magnonic devices with synergistic photonic and spin-electronic components on-chip. This technique operating at the telecommunication wavelength at 1550-nm has been recently developed ~\cite{yoon_prb2016, xiong2020detecting}. It takes advantage of the conventional microwave excitation of FMR and synchronizes such excitation with a GHz-modulated optical probe. A schematic illustration of the experimental setup is shown in Fig. \ref{MOD}a. The setup allows facile modulation at the GHz frequencies with both amplitude and phase controls and also phase-locking to a microwave source for FMR excitation. Figure \ref{MOD}b shows the intensity map of the FMR scan for the Fe/Pt bilayer. The optical signals with the phase information are obtained by the lock-in amplifier’s in-phase ($X$) and quadrature ($Y$) components. The total amplitude calculated by $\sqrt{X^2+Y^2}$ resembles the conventional microwave diode measurements and the extracted phase evolution is shown in Fig. \ref{MOD}c. 

Such a continuous-wave (CW) modulation capability makes this particular wavelength-band advantageous for studying magnetization dynamics in complex magnetic systems \cite{li2019simultaneous, xiong2020probing_magnon, li2019optical_detection, xiong2022tunable, inman2022hybrid}, such as quantum magnonic hybrids, patterned nanomagnets, spin ice, and 2D magnets. Compared to visible light wavelengths for magneto-optics, the method allows for the coherent tracking of gigahertz spin dynamics in a CW fashion, very much resembling a “lock-in” type of measurement that is commonly performed in many low-noise electric and spin transport measurements. In addition, the fiber-based optics allow for a compact integration with simultaneous electrical, thermal, and magnetic measurements with less susceptibility to typical mechanical vibrations, and therefore, having the advantage of being made into compact, tabletop or even portable systems with yet robust measurement performances for studying 2D magnets and related 2D heterostructures.

\subsection{Magnon-photon coupling probed in a superconducting resonator}

The interaction between microwave photons and magnons is determined by the mode volume overlap between the two. This essentially means that when we aim to probe spin dynamics of 2D vdW monolayers (which are typically $\mu$m sized flakes) using microwave photons, the photon mode must have a correspondingly small mode size to maximize the coupling strength and hence its sensitivity. One of the promising approaches to this is to use on-chip superconducting (SC) resonators ~\cite{Mandal_APL2020,Zhang_APL2021,Zollitsch_arXiv2022}. Strong coupling between photons in SC resonators and magnons in anti-ferromagnetic CrCl$_3$ bulk systems has been demonstrated ~\cite{Mandal_APL2020,Zhang_APL2021}.

Zollitsch {\it et al.} recently reported that it is possible to probe spin dynamics of Cr$_2$Ge$_2$Te$_6$ as thin as 11 nm using SC resonators ~\cite{Zollitsch_arXiv2022}. Cr$_2$Ge$_2$Te$_6$ flakes were exfoliated and transferred onto magnetically-active parts of high-quality-factor SC lumped element resonators with their small mode volume ($\approx$ 6000 $\mu$m$^3$), as schematically shown in Fig. \ref{MPC}a. The SC resonators are coupled to a microwave transmission line, through which the SC resonance modes are monitored. Figure \ref{MPC}b diplays a 2D color map of the microwave transmission coefficient ($S_\text{21}$) as a function of both external magnetic field and frequency. A clear SC photon mode is visible and the spectral distortion of the photon mode is found in a magnetic field range (590-620 mT) where the magnon Kittel mode is expected to have the same resonance frequency as the photon mode. As shown in Fig. \ref{MPC}c, this causes the observable linewidth broadening for the particular field range, which can be analysed by using a phenomenological model of coupled harmonic oscillators ~\cite{Zollitsch_arXiv2022}. The photon mode frequency ($\omega_\text{res}$) as well as the relaxation rate of the hybrid magnon-photon modes ($\kappa_\text{eff}$) were extracted by the spectral data as shown in Fig.~\ref{MPC}d and~\ref{MPC}e. $\kappa_\text{eff}$ is strongly enhanced due to the large relaxation rate of CGT magnons and using the plot shown in Fig.~\ref{MPC}e, the spin dynamics of the CGT magnon modes can be analysed (see details in ~\cite{Zollitsch_arXiv2022}). They also demonstrated the photon-magnon coupling using a 11 nm CGT flake (Fig.~\ref{MPC}f) where the CGT magnon modes are still detected through this measurement scheme. It turns out that CGT magnon modes have a large dissipation rate, posing a fundamental bottleneck for achieving the strong coupling at this limit.  However,  multiple avenues for improving the coupling strength are available to overcome this challenge. Such as by introducing nm-scale constrictions with resonator mode volume reduction ~\cite{McKenzie2019, Gimeno2020}, and mode overlap enlargement using large-scale 2D vdW material transfer techniques ~\cite{Huang2020}.  With the successful creation of coherent photon-magnon hybrids (magnon-polaritons),  they can be used for efficient energy-transfer between polaritonic states with different degrees of freedom in vdW systems ~\cite{Basov_Science2016,Low_Nature2017}. 

\begin{figure*}[ht]
 \centering
 \includegraphics[width=1.0\textwidth]{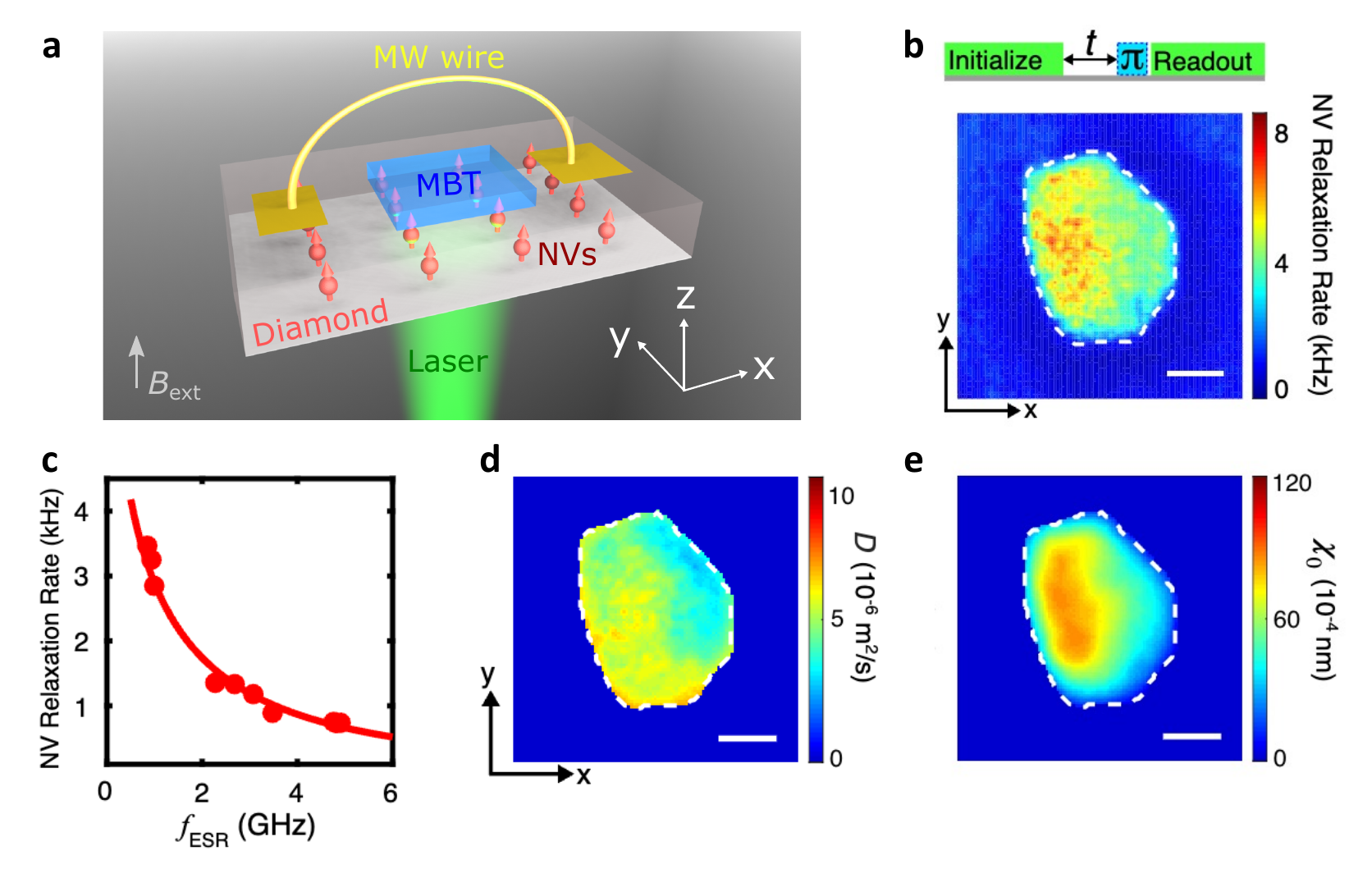}
 \caption{\textbf{a.} Schematic of an NV wide-field magnetometry measurement platform. \textbf{b.} NV spin relaxation map measured for an exfoliated MnBi$_4$Te$_7$ (MBT) flake at ESR frequency \textit{f}$_\mathrm{ESR}$ of 1.0 GHz at 8.5 K. The boundary of the MBT flake is outlined with white dashed lines. \textbf{c.} Spatially averaged NV spin relaxation rates measured on the MnBi$_4$Te$_7$ flake (red dots) as a function of \textit{f}$_\mathrm{ESR}$ at 8.5 K. The experimental results are in excellent agreement with the theoretical prediction (red curve). \textbf{d-e.} 2D images of spin diffusion constant (D) and magnetic susceptibility $\chi_0$ of the MBT flake at 8.5 K. The scale bar is 4 $\mu$m. All panels are adapted from Ref. ~\cite{mclaughlin2022quantum}.}
 \label{NV}
\end{figure*}

\subsection{Nitrogen-vacancy electron spin resonance} 

Advanced quantum sensing techniques provide another perspective to investigate the rich physics of 2D magnets. Nitrogen-vacancy (NV) centers, optically active spin defects in diamond, are a prominent candidate in this category~\cite{degen2017quantum}. Over the past years, NV magnetometry techniques have been demonstrated as a transformative tool in exploring the local static and dynamic spin behaviors in 2D magnets with competitive field sensitivity and spatial resolution. Examples include nanoscale imaging layer-dependent 2D magnetization~\cite{thiel2019probing}, 2D magnetic domains~\cite{sun2021magnetic}, magnetization reversal processes~\cite{broadway2020imaging}, moiré magnetism~\cite{song2021direct}, room-temperature 2D ferromagnetism~\cite{fabre2021characterization,chen2022revealing}, and others~\cite{zhang2021ac,mclaughlin2022quantum}. Here, we briefly discuss the opportunity to use NV relaxometry method to probe intrinsic spin fluctuations in 2D magnetic materials, which is challenging to access by conventional magnetometry techniques.

Figure \ref{NV}a shows a schematic of NV-based wide-field imaging platform. A \ce{MnBi4Te7} nanoflake is transferred onto the diamond surface with shallowly implanted NV ensembles ~\cite{mclaughlin2022quantum}. Figure \ref{NV}b show an example of the 2D NV relaxation maps. NV relaxometry measurements take advantage of the dipolar interaction between spin density of a magnetic sample and proximate spin sensors~\cite{finco2021imaging,du2017control,flebus2018quantum,mclaughlin2022quantum}. At thermal equilibrium, fluctuations of the longitudinal spin density will generate low-frequency fluctuating magnetic fields at the NV electron spin resonance frequencies \textit{f}$_\mathrm{ESR}$, resulting in enhancement of NV relaxation rates. By measuring the NV relaxation rate as a function of \textit{f}$_\mathrm{ESR}$ and fitting it into a theoretical model (Fig. \ref{NV}c), we can extract key material parameters such as local spin diffusion constant ($D$) and magnetic susceptibility ($\chi_0$) of the \ce{MnBi4Te7} flake as shown in Fig. \ref{NV}d and \ref{NV}e ~\cite{mclaughlin2022quantum}. Temperature and magnetic field dependent NV relaxation rate measurements can also be used to reveal the magnetic phase transitions of 2D magnets. It is instructive to note that the spatial resolution of the NV wide-field imaging techniques typically stays in the hundreds of nanometers regime, which is fundamentally set by the optical diffraction limit~\cite{broadway2020imaging}. For scanning NV microscopy, the spatial sensitivity can ultimately reach the regime of tens of nanometers, offering an attractive platform to reveal the detailed microscopic spin textures and dynamic responses in various 2D quantum materials~\cite{maletinsky2012robust,pelliccione2016scanned}.

\subsection{Other FMR techniques}

\begin{figure*}[ht!]
 \centering
 \includegraphics[width=0.98\textwidth]{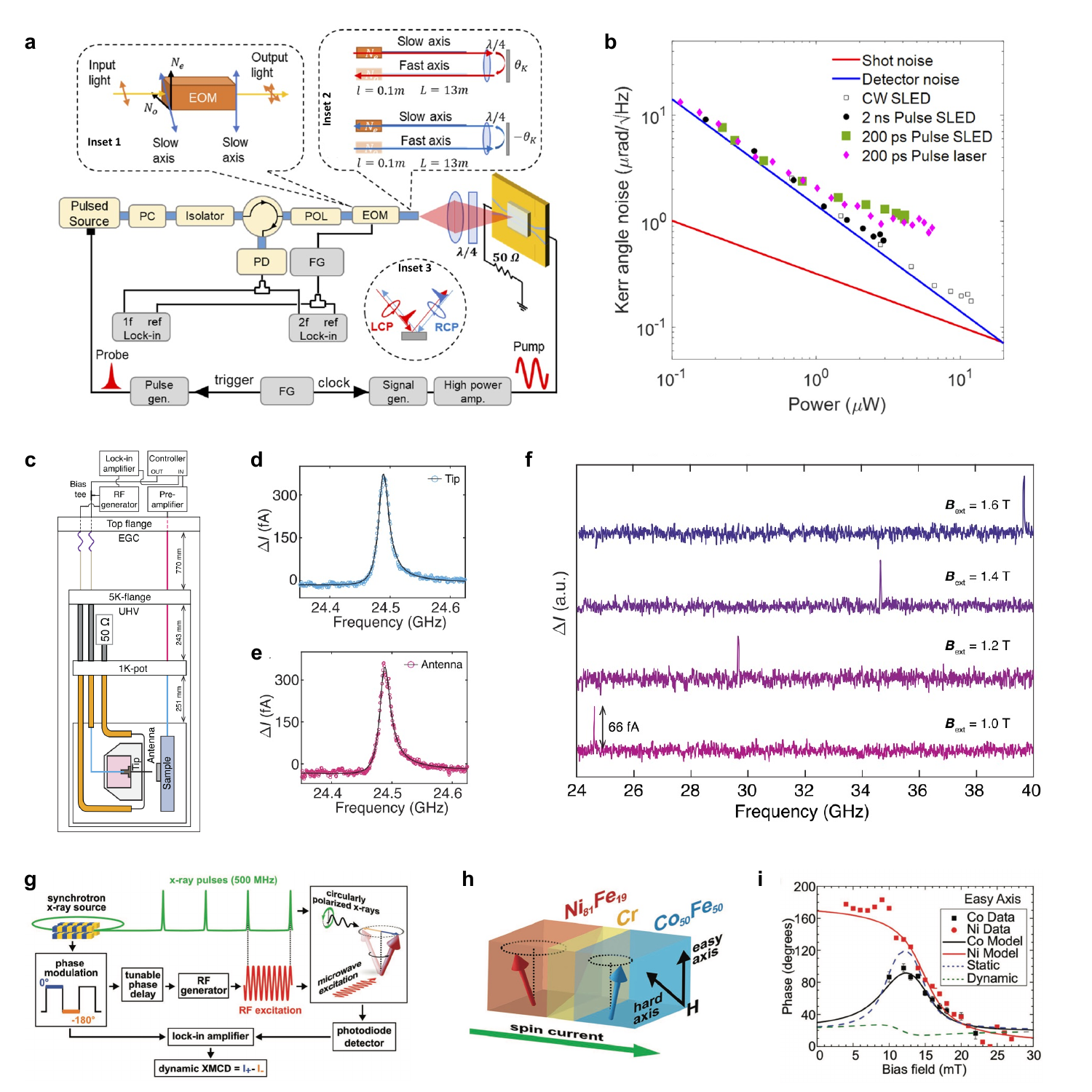}
 \caption{\textbf{a.} Schematic of the time-resolved zero-area Sagnac interferometer (TR-ZASI). PC: polarization controller, POL: polarizer, PD: photoreceiver, and FG: function generator. \textbf{Inset 1}: The polarization axis of the electro-optic modulator (EOM). \textbf{Inset 2}: The two counterpropagating paths in the TR-ZASI. \textbf{Inset 3}: Optical pulses near the sample. LCP (RCP) means left (right) circular polarization. \textbf{b.} Noise characteristics of TR-ZASI for varying light sources. \textbf{c.} Wiring scheme for electron spin resonance scanning tunneling microscopy (ESR-STM). Two methods are adopted to apply RF voltages to the tunnel junction: STM tip and RF antenna. \textbf{d-e.} Comparison of ESR spectra measured using the tip and the antenna for RF transmission. \textbf{f.} ESR spectra of hydrogenated titanium (TiH) atoms at varying magnetic fields measured using the antenna. \textbf{g.} Synchrotron-based x-ray detected FMR setup. Microwaves are generated at a frequency of the higher harmonic of the 500 MHz storage ring frequency to ensure a fixed phase relationship between the RF pump field and the probing x-ray pulses. The phase of the RF excitation is modulated by 180° to probe the change in magnetization between opposite sides of the cone of precession. \textbf{h.} Architecture of \ce{Ni81Fe19}/Cr/\ce{Fe50Co50} spin valve device to generate and detect spin current. \textbf{i.} Phase of the \ce{Fe50Co50} spin precession in proximity to the FMR condition of \ce{Ni81Fe19} with field applied along the easy axis of \ce{Fe50Co50}. Panels \textbf{a-b} are adapted from Ref. ~\cite{heo2022sagnac}. Panels \textbf{c-f} are adapted from Ref. ~\cite{hwang2022development}. Panels \textbf{g-i} are adapted from Ref. ~\cite{klewe2020element}.}
 \label{Other}
\end{figure*}

It is worth noting that 1550 nm CW laser has also been used in the all-fiber design of Sagnac interferometer for magneto-optical measurements of ferromagnetic resonance ~\cite{liu2019observationEP}. Sagnac interferometers have proven as a powerful technique that can measure the magneto-optical Kerr effect with 100 nrad/$\sqrt{\mathrm{Hz}}$ sensitivity using only a 10 $\mu$W optical power without the magnetic field modulation ~\cite{spielman1990test, spielman1992measurement, xia2006modified, xia2006high, fried2014scanning, zhu2017symmetry, zhu2021oblique}. Recently, this technique is employed in the magneto-optical detection of photoinduced magnetism via chirality-induced spin selectivity in 2D chiral hybrid organic-inorganic perovskites ~\cite{huang2020magneto}. Moreover, by replacing the CW 1550 nm laser to a pulsed one, Heo \textit{et al} develops a time-resolved zero-area Sagnac interferometer (TR-ZASI) for magneto-optical measurements (see Fig.~\ref{Other}a) ~\cite{heo2022sagnac}. As shown in Fig.~\ref{Other}b, in the pulse mode, similar to the case of continuous wave (CW), each pulse will interfere after propagating the clockwise and counterclockwise loops. Also, it has been demonstrated that the pulse mode and the CW mode give the same calibrated Kerr angle. This technique is then used to measure time-resolved Kerr signal at the FMR of a Permalloy film, and the results are fully consistent with other techniques, such as vector network analyzer. Temporal resolution of hundreds of picoseconds is achieved, maintaining the advantages of the Sagnac interferometer. The analysis of the noise (see Fig.~\ref{Other}c) shows that the TR-ZASI can achieve 1 $\mu$rad/$\sqrt{\mathrm{Hz}}$ sensitivity at a 3 $\mu$W optical power in the pulse mode. This technique is expected to contribute to magneto-optical measurements of various fast dynamics in ps and ns ranges.

In recent years, the combination of electron spin resonance and scanning tunneling microscopy (ESR-STM) has been demonstrated as a technique to detect magnetic properties of single atoms on surfaces and to achieve sub-microelectronvolts energy resolution ~\cite{baumann2015electron}. In an ESR-STM experiment performed in Fe atoms absorbed on MgO thin films, it has been shown that single-atom electron spin resonance properties can be tuned by combing a vector magnet and the field from the spin-polarized STM tip ~\cite{willke2019tuning}. Moreover, by the mixing of a continuous RF voltage to the STM junction, an rf spin-polarized tunneling current is generated from the magnetic tip, which drives a coherent magnetic precession in a ferromagnetic thin film. Hwang \textit{et al} report a homebuilt ESR-STM incorporated with a Joule-Thomson refrigerator and a two-axis vector magnet ~\cite{hwang2022development}. In addition to the early design of wiring to the STM tip, they apply RF voltages using an antenna (see Fig.~\ref{Other}d). Direct comparisons of the ESR spectra measured using these two methods (see Fig.~\ref{Other}e and ~\ref{Other}f) shows consistent intensity, lineshape, and resonance frequency, indicating that their mechanisms of ESR driving and detection are the same. As shown in Fig.~\ref{Other}g, the antenna method is employed to measure ESR spectra of hydrogenated titanium (TiH) atoms on Mgo/Ag(100) at different magnetic fields. This technique permits the study of nanoscale magnetic systems and magnetic skyrmions ~\cite{herve2019towards}, bearing great potential for quantum sensing and coherent manipulation of quantum information.

Finally, x-ray pulses from synchrotron radiation enable element, site, and valence state resolution of magnetization and spin dynamics ~\cite{goulon2007element, arena2009compact}. Therefore, the technique of x-ray detected FMR (XFMR) combines FMR and x-ray magnetic circular dichroism (XMCD), in which the sample is pumped by an RF magnetic field to generate a precession of the magnetization, which is then probed using the XMCD effect with magnetic and chemical contrast. XFMR experiments can be carried out in time-averaged and time-resolved manners. In particular, time-resolved measurements are performed in a transverse geometry, where the magnetization is oriented perpendicular to the incident x-ray pulses ~\cite{bailey2013detection}. In this geometry, the magnetic moments are continuously excited by the RF field while the response from the phase-dependent magnetization components along the x-ray beam is probed stroboscopically. Time-resolved XFMR can thus be used to measure both amplitude and phase of the spin precession. As shown in Fig.~\ref{Other}h, microwaves are generated at a frequency of the higher harmonic of the storage ring frequency (500 MHz) to ensure a fixed phase relationship between the RF pump field and the probing x-ray pulse. By incrementally delaying the phase of the RF field with respect to the timing of the x-ray pulses, the complete spin precession cycle can be mapped. The phase and chemical resolution of XFMR allow to determine the contribution of interlayer coupling and dynamic coupling of each layer in multilayer samples. As shown in Fig.~\ref{Other}i and~\ref{Other}j, it has unique capabilities to probe spin-transfer torque and spin currents in spin valve devices ~\cite{klewe2020element, van2017time}. These advantages can be applied to study magnetic vdW heterostructures and vdW magnetic tunneling junctions.

\begin{table*}[hb!]

\textbf{\caption{Magnetic properties of emerging vdW magnets and prototypical bulk magnets.}}
\vspace{5pt}
\centering
\renewcommand{\arraystretch}{1.7}
\begin{tabular}{|p{2.4cm}||p{1.5cm}|p{1.5cm}|p{2.0cm}|p{2.0cm}|p{1.7cm}|p{3.0cm}|p{1.7cm}|}
\hline
\textbf{Material}  & \textbf{$g_{\perp}$}     &   \textbf{$g_{\parallel}$}   &  \textbf{$H_k$ (kG)}     & \textbf{$\alpha$}   & \textbf{$T_c$ (K)} & \textbf{Synthesis method} & \textbf{Reference}  \\
\hline
CrCl$_3$  & 1.970 \newline (100 K)  & 1.990 \newline (100 K)  & $\sim$0 & - & 17(bulk) &  CVT & ~\cite{PhysRevMaterials.4.064406}\\
CrI$_3$              & 2.07 \newline (1.5 K)   & -  & 28.6 (1.5 K)  & -  & 70 (bulk) \newline 45 (ML) & CVT, MBE  & ~\cite{dillon1965magnetization} \\
CrBr$_3$             & -  & -  & 5.6 (10 K) & 0.009 \newline (10 K) &  47 & CVT, flux &  ~\cite{Chunli_Ref.7shen2021multi}   \\
Fe$_5$GeTe$_2$            & 1.99 \newline (300 K) & 2.21 \newline (300 K)  & 1.8 (300 K)  & 0.035 \newline (300 K) & 300 & CVT, CVD, MBE & ~\cite{alahmed2021magnetism}\\
CrSiTe$_3$   & 2.70 \newline (34 K)  & 1.75 \newline (34 K)  &  20 (2 K)  &  -   & 34.15  & flux & ~\cite{li2022anomalous}\\

Cr$_2$Ge$_2$Te$_6$      & 2.10 \newline (2 K) &  2.18 \newline (2 K) & 5.82 (2 K)  & 0.01-0.08 \newline (10 K)  & 64.7 & CVT, flux & ~\cite{zhang2020laser}    \\
Fe$_3$GeTe$_2$          & - & - & 12 (185 K) & 0.58 \newline (185 K)   & 150-220 & CVT, MBE & ~\cite{ni2021magnetic} \\

Y$_3$Fe$_5$O$_{12}$ (YIG)   & 2 (300 K) & 2 (300 K) & 0.09-0.4 \newline (300 K) & 8.58 $\times 10^{-5}$ \newline (300 K) & 567 & Sputtering, PLD & ~\cite{chang2014nanometer} \\

NiFe         & 2 (300 K) & 2 (300 K) & 0        & 0.005-0.008 (300 K) & 553 (bcc) \newline 872 (fcc) & Sputtering, MBE & ~\cite{yu2008curie} \\
\hline
\end{tabular}

\begin{tablenotes}
    \item[1] CVT: Chemical vapor transport; MBE: molecular beam epitaxy; CVD: Chemical vapor deposition; \newline PLD: Pulsed laser deposition.
  \end{tablenotes}
\end{table*}

\section{Conclusion and outlook}

Advances in the 2D vdW magnet exploration have spurred new interests in spintronics. As we have discussed throughout the text, probing spin dynamics provides access to key parameters such as the exchange interaction, magnetic anisotropy, $g$-factor, spin-wave eigenmodes, and spin-orbit torque in 2D vdW magnets and heterostructures, which is crucial for designing ultrafast spintronic devices. The magnetic properties including $g$-factors extracted from FMR along $H\parallel c$ ($g_{\perp}$) and $H\parallel ab$ ($g_{\parallel}$), effective magnetization ($H_k$), damping rate ($\alpha$), transition temperature ($T_c$), and synthesis method for emerging vdW magnets and prototypical bulk magnets (yttrium iron garnet, or YIG) and magnetic thin films (NiFe) are summarized in \textbf{Table III}.

Benefited from the rich library of 2D vdW materials with their distinct properties in optics, magnetism, superconductivity, etc, the various synthetic vdW systems offer a great material playground for novel hybrid quantum systems, where coherent signal and energy transduction is often demanded. Over the last few years, 2D vdW compounds have emerged as promising contenders for hybridized polariton sciences, demonstrating efficient, coherent couplings between distinct material excitations, such as photon, phonon, and exciton. In addition, as a rapidly growing subfield of quantum engineering, magnonic (dealing with spin dynamics, i.e. magnons) have provided augmented engineering capabilities in coherent information processing ~\cite{Awschalom_IEEE2021}. Understanding and engineering the spin dynamics in vdW systems will lead to new opportunities in hybrid quantum transduction using magnons, where the ease of 2D layer exfoliating and stacking and the magnon advantages (e.g. strong coupling with other excitations) are combined. One of the main challenges in the full integration of 2D magnets in quantum technologies (e.g. computing, photodetectors, information storage, processors, qubits, on-chip platforms) relies on the finding of truly scalable room-temperature magnetic layers where quantum effects are substantial. Currently, there is no clear recipe where to find them in the available library of 2D compounds. What we know at the moment is that FMR provides an intimate dialogue with the fundamental interactions present in the spin dynamics at each system. From them, we can estimate whether new routes can be used to enhance such building blocks, for instance, in terms of different assemblies and/or chemical modifications.

Using vdW assembly such as moiré lattice, one can combine different magnetic materials to induce magnon-magnon coupling without out-of-plane field by breaking sublattice exchange symmetry.  In addition, recent simulation works suggest that synthetic vdW magnets exhibit both dipolar coupling and Ruderman–Kittel–Kasuya–Yosida (RKKY) exchange coupling among multiple vdW sublattices, which open up new opportunities for very strong magnon hybridization~\cite{ZhangCrCl3,ZhangEP}. These intriguing physics and promising applications will attract great interest in ferromagnetic resonance studies of the microscopic quantities associated with spin dynamics and magnon hybridization. Moreover, electric field can be used to modulate the resonance frequency, damping and coupling strength in vdW magnet-based magnonic devices. A recent theoretical work predicts electric field controlled magnonic dynamics at exceptional points in a parity–time symmetric waveguide \cite{YIGfilmEPNatComm}. ST-FMR thus provides a powerful tool to investigate and realize such spin-orbit interaction driven interfacial phenomena.

\section{Acknowledgements}

W.J. acknowledges support by NSF EPM Grant No. DMR-2129879 and Auburn University Intramural Grants Program. C.T. acknowledges financial support from the Alabama Graduate Research Scholars Program (GRSP) funded through the Alabama Commission for Higher Education and administered by the Alabama EPSCoR. Y.X., J.I., and W.Z. acknowledge partial support from U.S. National Science Foundation under Grants No. ECCS-2246254 for the preparation of this manuscript.  N.J.M. and C.R.D. were supported by the Air Force Office of Scientific Research under Award FA9550-20-1-0319. EJGS acknowledges computational resources through CIRRUS Tier-2 HPC Service (ec131 Cirrus Project) at EPCC (http://www.cirrus.ac.uk) funded by the University of Edinburgh and EPSRC (EP/P020267/1); ARCHER UK National Supercomputing Service (http://www.archer.ac.uk) {via} Project d429, and the UKCP consortium (Project e89) funded by EPSRC grant ref EP/P022561/1. EJGS acknowledge the Spanish Ministry of Science's grant program ``Europa-Excelencia'' under grant number EUR2020-112238, the EPSRC Open Fellowship (EP/T021578/1), and the Edinburgh-Rice Strategic Collaboration Awards for funding support.  

% \bibliography{Main}% Produces the bibliography via BibTeX.
\bibliographystyle{apsrev4-1}
\nocite{apsrev41Control}
\bibliography{Main.bib}% Produces the bibliography via BibTeX.

\end{document}